\documentclass[prd,aps,groupedaddress,eqsecnum,twocolumn,floats,superscriptaddress,preprintnumbers]{revtex4}
\usepackage{amsfonts,amsmath,epsfig,amssymb,latexsym,eucal,color,graphicx,array,subfigure,bm}
\usepackage{eucal,times}
\usepackage{graphicx,wrapfig}

\begin{document}


\mbox{}\\[10em]
\title{Attractor Universe in the Scalar-Tensor Theory of Gravitation}

\author{Kei-ichi {\sc Maeda}}
\email{maeda@waseda.jp}

\address{
Department of Physics, Waseda University,
Okubo 3-4-1, Shinjuku, Tokyo 169-8555, Japan}

\address{
 Advanced Research Institute for Science and Engineering,
Waseda  University, Shinjuku, Tokyo 169-8555, Japan}

\address{
Waseda Institute for Astrophysics, Waseda University,
Shinjuku,  Tokyo 169-8555, Japan
}

\author{Yasunori {\sc Fujii}}
\email{fujii@gravity.phys.waseda.ac.jp}

\address{
 Advanced Research Institute for Science and Engineering,
Waseda  University, Shinjuku, Tokyo 169-8555, Japan}

\date{\today}


\begin{abstract}
In the scalar-tensor theory of gravitation it seems nontrivial to
 establish if solutions of the cosmological equations in the presence of
 a cosmological constant behave as
 attractors independently of the initial values.  We develop a general
 formulation in terms of  two-dimensional phase space.  
We show that there are two kinds of fixed points, 
one of which is an attractor depending on the coupling constant
and equation of state.
 In the case with a power-law potential in the Jordan frame, 
we also find new type of inflation caused
by the coupling to the matter fluid. 
\end{abstract}
\maketitle

\section{Introduction}

Einstein's General Relativity has proved to be the simplest theory for 
successful understanding of a number of experiments and observations.  Still, 
on the other hand, there seem to be growing indications that a 
yet-to-be-discovered scalar field might play fundamental roles in cosmology.  
A list of the possible sources include the one expected from higher-dimensional
 theories of gravitation such as superstring/M theory and a scalar field 
(a volume modulus) which  couples to 4-dimensional gravity after
 compactification, also those in the context of the brane world scenario.
  We may reasonably expect that realistic consequences of these hypotheses can
 be implemented in terms of the scalar-tensor theory proposed first by
 Jordan \cite{jordan}, developed later by Brans and Dicke \cite{bd}.

One of the recent focuses of the scalar field is aimed particularly at the 
origin of the dark energy which appears to be required from the observed
 acceleration of the universe, with the renewed interest of today's
version of  the cosmological constant problem, culminating to the twin
questions;  fine-tuning problem and the coincidence problem.  We argued
\cite{Fujii_Maeda,yf82,ptpinv,antip} 
that the scalar-tensor theory is precisely that
causes the behavior,  $\Lambda \sim t^{-2}$ 
realized numerically by $10^{-120}\sim
(10^{60})^{-2}$
\cite{footnote1}, where $10^{60}$ is today's age  of the universe in
units of the Planck time $\sim 10^{-43}{\rm s}$, hence preparing another
simple implementation of the scenario of the decaying 
cosmological constant as discussed in \cite{dolgov} and  \cite{freese}.

Before reaching  conclusions to be compared with observation, however, we must 
go through certain complications including details on the choice of the 
conformal frames among other things.   Also many consequences derive from 
leaving the Brans-Dicke model \cite{bd}, as required ultimately by a single technical 
aspect on the attractor nature of the cosmological solutions.  This motivated 
us to develop a general framework of studying dynamics of the system including 
the scalar field taking the unique roles of the conformal transformation 
properly into account. 

After a brief introduction of the action of the scalar-tensor theory in both
 conformal frames, the Jordan and the Einstein frames in Section II, we enter 
Section III to develop a formulation in the Einstein frame in which we
 may trace  how the cosmological solutions evolve in 
 two-dimensional phase space. 
 We assume the presence of the exponential potential of the scalar field 
corresponding to the simple cosmological constant in the Jordan frame.  It is 
crucially important to make a right choice of the new extended time-coordinate
 other than the conventional cosmic time.  Also of central importance is to 
deal with self-autonomous systems.

In Subsection III-A we discuss the fixed points in phase space. 
We show that there are two different sets of fixed points: one (FP1)
is for the well-known universe of scalar-field 
dominance, and the other (FP2)
 represents a new type of the universe in which the matter fluid energy is scaled to the potential of the scalar field
because of the coupling with the matter fluid.
We  find that the universe of the latter type always 
expands in the same way as  the radiation-dominant
universe in the Einstein frame.
In Subsection III-B,  we present the stability analysis and the 
behavior of the attractor solutions.  Flows of trajectories in phase space are 
illustrated for the examples taken from the radiation-dominant 
universe. 
Basically the same analysis will be repeated in Section IV now in the Jordan 
frame.

 In Section V we generalize the argument to the power-law potential with
 a monomial of the scalar field multiplied by the cosmological constant chosen 
in the preceding section in the Jordan frame.  
We find a new type of inflation at the fixed point FP2
even for the potential which is too steep to cause inflation 
without the coupling with the matter fluid. 
Section VI is devoted to the  concluding remarks.

In Appendix A, we discuss the effect of the curvature term not included in
the preceding sections.  In the subsequent three appendices, we add
related discussions on the accelerating universe.  Appendix B reveals the
presence of a complication in the numerical analysis in the Jordan frame,
while in Appendix C we discuss the attractor nature of the
scale-invariant model as an 
alternative to the Brans-Dicke model.  The final Appendix D will be
devoted to offering another simplified approach to the power-law potential.

\section{scalar-tensor theory:
Jordan frame vs Einstein frame}
We discuss cosmology in the scalar-tensor theory of gravitation.
We assume the presence of a potential $V(\phi)$
 in the Jordan frame.
The action is
\begin{eqnarray}
S&=&S_g+S_m
\,,
\label{action_Jordan}
\end{eqnarray}
where 
\begin{eqnarray}
S_g&=&\int d^4x \sqrt{-g}\left[{\xi \over 2} 
\phi^2 R-{\epsilon\over 2}(\nabla\phi)^2-V(\phi)
\right]
\label{gravity_action_Jordan}
 \\
S_m&=
&
\int d^4x \sqrt{-g}\,L_{\rm m}(g, \phi)
\,,
\label{matter_action_Jordan}
\end{eqnarray}
with $\epsilon=\pm 1$ \cite{footnote2}.     
 Note that the matter action $S_m$ is assumed to have
no scalar field $\phi$ according to the Brans-Dicke model \cite{bd}, in
which Weak Equivalence Principle (WEP) is intended to be respected.  
The scalar-tensor theory defined in this way is equivalent to  the traditional Jordan-Brans-Dicke theory with an added potential $U(\varphi)=
V(\phi)$, often expressed as
\begin{eqnarray}
S_g&=&\int d^4x \sqrt{-g}\left[ 
\varphi R-{\omega\over \varphi}(\nabla\varphi)^2-U(\varphi)
\right]
\,,
\label{action_BD}
\end{eqnarray}
with the Brans-Dicke constant $\omega=\epsilon/(4\xi)$ and $\varphi
=(\xi /2)\phi^2$.

Although we may further extend this type of the scalar-tensor theory
 with an arbitrary function of the
scalar field $\phi$ multiplied with $R$, we confine ourselves to
the original simple $\varphi$, because it features global
scale invariance except generally for the $V(\phi)$ term.

The choice $\epsilon=-1$ in (\ref{gravity_action_Jordan}) is closely related to
string theory.  The $D$-dimensional 
action for the zero-modes in the closed
string sector is given \cite{Callan} by 
\begin{eqnarray}
S=\frac{1}{2}\int d^Dx \sqrt{-g} e^{-2\Phi}\left[R(g)+4\left(
\nabla\Phi\right)^2\right]
\,, \label{string}
\end{eqnarray}
which, re-expressed according to our own sign convention as in (1.30) of 
\cite{Fujii_Maeda}, corresponds to the first two terms in (\ref{gravity_action_Jordan}) with  $\epsilon=-1$ and
$\xi=1/4$,  hence $\omega=-1$, by introducing $\phi=2e^{-\Phi}$.

We can always move to the Einstein frame by a conformal transformation
\cite{conformal_trans1,conformal_trans2,conformal_trans3}
\begin{eqnarray}
g_{\mu\nu}\rightarrow g_{*\mu\nu}=\Omega^2 g_{\mu\nu} 
\label{cftmetric}
\,
\end{eqnarray}
where
\begin{eqnarray}
\Omega^2 =\xi \phi^2 =\exp (2\zeta\sigma )
\,,
\label{Omega}
\end{eqnarray}
with
\begin{eqnarray}
\zeta^{2} \equiv \left( 6+\epsilon \xi^{-1}\right)^{-1} =\left(
		  6+4\omega \right)^{-1}
\,,
\label{zeta}
\end{eqnarray}
which defines a canonical scalar field $\sigma$ in the Einstein frame;
\begin{eqnarray}
S&=&\int d^4x \sqrt{-g_*}
\left[{1 \over 2} R_*-{1\over 2}(\nabla_{\hspace{-.2em}*}
\sigma)^2-V_*(\sigma)
\right]
\nonumber \\
&
+
&\int d^4x \sqrt{-g_*}L_{\rm m}( \psi_*, g_*, \sigma)
\,,
\label{Eaction}
\end{eqnarray}
where
\begin{eqnarray}
V_*(\sigma)=\exp\left(-4\zeta\sigma\right) V(\phi)
\,. \label{vstar}
\end{eqnarray}
We mark the quantities in the Einstein frame with the subscript $*$,
while those in the Jordan frame are left unmarked, unless otherwise
indicated.  This is in accordance with the notation used  in \cite{Fujii_Maeda}.

We point out that $\epsilon =-1$,  apparently indicating a ghost nature of the
non-diagonalized field $\phi$, is a real difficulty only if 
$\zeta^{2}$ 
turns out to be negative implying a negative energy for the diagonalized field
$\sigma$.  
We always assume  the condition
\begin{eqnarray}
\zeta^{2}>0
\,.
\end{eqnarray}
 This can be obeyed even if $\epsilon =-1$ if
$\xi >1/6$. 
 Imposing $\xi >0$,
which we assume throughout this paper, due to the required positivity of the energy of tensor gravity, 
we find that  $\epsilon=1$ allows any $\xi$ but with $\zeta^2 <1/6$,
while $\epsilon = -1$ constrains $\xi>1/6$ and $\zeta^2 >1/6 $, as
displayed graphically in Fig. 1 of \cite{ptpinv}.

The parameters in (\ref{string}) gives $\zeta^2
=(D-2)/4$ which is $1/2$ for $D=4$.  If $\epsilon=-1$ and $\xi=1/6$,
 we find $\zeta^2\rightarrow \infty$, implying  no kinetic term in the
Einstein frame, hence no degree of freedom.  We do not consider this
choice any further.

Because we assume that no
$\phi$ field enters $L_m$ in the Jordan frame, we find that the 
energy-momentum of the matter fluid is conserved in the Jordan frame;
\begin{eqnarray}
\nabla^\nu  T^{\mu}_{~\,\nu} =0
\label{em_conservation_Jordan}
\, ,
\end{eqnarray}
for which WEP is respected.
The 
energy-momentum tensor in the Einstein frame is obtained by
\begin{eqnarray}
T^{\,\mu}_{\hspace{-.2em} *\hspace{.2em}\nu} =\exp\left(-4\zeta\sigma\right)
T^{\mu}_{\hspace{.5em}\nu}
\,,
\label{cnfemtensor}
\end{eqnarray}
which is no longer conserved;
\begin{eqnarray}
\nabla_{\hspace{-.2em}*}^\nu T^{\mu}_{*\hspace{.2em}\nu}  =-\zeta T_* 
\nabla_{\hspace{-.2em}*}^\mu\sigma, 
\label{em_conservation_Einstein}
\,,
\end{eqnarray}
where $T_*=T^{\,\rho}_{*\hspace{.2em}\rho}$.  Note that the universal free-fall
(UFF) is still maintained, as an expression of WEP.


\section{Cosmology with a cosmological
constant  : Analysis in the Einstein frame}
We discuss cosmology in the scalar-tensor theory
with a cosmological constant $V=V_0$.
In this section, we discuss it in the Einstein frame, though the
analysis in the Jordan frame is given in the next section.

\subsection{The Basic Equations and the Fixed Points}

The metric of isotropic and homogeneous universe
is given by the FRW form:
\begin{eqnarray}
ds_*^2=-dt_*^2+a_*^2 ds_3^2
\,,
\label{lineE}
\end{eqnarray}
where $ds_3^2$ is the metric of
maximally symmetric three-dimensional space with 
the curvature constant $k=0$ or $\pm 1$.

The basic equations in the Einstein frame are
\begin{eqnarray}
&&
H_*^2+{k\over a_*^2}={1\over 3}\left(
{1\over 2}\dot{\sigma}^2+V_*+\rho_*
\right), \label{fundeq1_1}
\\
&&
\ddot{\sigma}+3H_*\dot{\sigma}
+{\partial V_*\over \partial \sigma}
=\zeta(\rho_*-3 P_*),
\label{fundeq1_2}
\end{eqnarray}
where $H_*=\dot{a}_*/a$, $P_*$,
 and $\rho_*$ 
are the Hubble expansion parameter,
the pressure, and the energy density in the Einstein frame,
respectively.
The dot implies $d/dt_*$ throughout in the Einstein frame.  
Eq. (\ref{em_conservation_Einstein}) in the
Einstein frame is then re-expressed as 
\begin{eqnarray}
\dot{\rho}_*+3H_*(P_*+\rho_*)=
- \zeta \dot{\sigma}(\rho_*-3P_*) 
\label{energy_conservation_Einstein}
\,.
\end{eqnarray}

Assuming the equation of state $P_*=(\gamma-1)\rho_*$, we further re-express
Eqs. (\ref{fundeq1_1})-(\ref{energy_conservation_Einstein}) into

\begin{eqnarray}
&&
H_*^2+{k\over a_*^2}={1\over 3}\left(
{1\over 2}\dot{\sigma}^2+V_*
+\rho_*
\right), \label{fundeq2_1}
\\
&& \ddot{\sigma}+3H_*\dot{\sigma}
-4\zeta V_*
=\zeta(4-3\gamma) \rho_*,\label{fundeq2_2} 
~~~~~~~
\\
&&
\dot{\rho}_*+3\gamma H_* \rho_*=
-\zeta (4-3\gamma)\dot{\sigma} \rho_* .
\label{fundeq2_3}
\end{eqnarray}

We now introduce a  new dimension-free time coordinate $\tau_*$
by
\begin{eqnarray}
d\tau_*= 2 \sqrt{V_*}dt_*
\,. 
\label{tau1}
\end{eqnarray}
 We further introduce 
$
{\cal H}_*={a'_* / a_*} \, ,
$
where the prime is for a differentiation with respect to $\tau_*$.  We
 then put Eqs. (\ref{fundeq2_1})-(\ref{fundeq2_3}) into the new
 form
\begin{eqnarray}
&&
{\cal H}_*^2+ {k\over 4 V_*a_*^2}
={1\over 6}\left[{\sigma}'^2+{1\over 2}
\left(1
+ {\rho_* \over  V_*}\right)
\right] \, ,
\label{Hamiltonian1}
\\
&&
{\sigma}''
+3{\cal H}_*{\sigma}'
-\zeta\left[2{\sigma}'^2
+1+{(4-3\gamma)\rho_* \over 4 V_*}\right]=0 \,,
~~~~~~~
\label{sigma1}
\\
&&
\rho'_* +3\gamma {\cal H}_* \rho_*=
-\zeta(4-3\gamma ){\sigma}' \rho_* 
\label{em_cons1}
\,.
\end{eqnarray}

Focusing on $k=0$, 
we  differentiate Eq. (\ref{Hamiltonian1}) 
with respect to
$\tau_*$, to obtain
\begin{eqnarray}
{\cal H}'_*
=
-{2-\gamma\over 4}{\sigma}'^2
+{\gamma\over 8}
+2\zeta{\sigma}'{\cal H}_*
-{3\gamma \over 2}{\cal H}_*^2
\label{Einstein2}
\,.
\end{eqnarray}
Here we have used Eqs. (\ref{Hamiltonian1}) and (\ref{sigma1}) 
as well as the equation 
\begin{eqnarray}
(\rho_*/V_*)'=-3\gamma( {\cal H}_*  -\zeta {\sigma}')
(\rho_*/V_*) 
\label{em_cons2}
\,,
\end{eqnarray}
which is obtained from Eq. (\ref{em_cons1}) and the definition
(\ref{vstar}) of $V_*$.

In the same way we put  Eq. (\ref{sigma1}) into
\begin{eqnarray}
{\sigma}''=-3{\cal H}_*{\sigma}'
+{3\gamma \zeta\over 4}\left( 2 {\sigma}'^2
+1\right)+3\zeta (4-3\gamma){\cal H}_*^2
\label{sigma2}
\,,~~~~
\end{eqnarray}
\mbox{}\\[.8em]
where Eq. 
(\ref{Hamiltonian1}) has been used to obtain the last term on the
 right-hand side.
A set of equations (\ref{Einstein2}) and  (\ref{sigma2})  gives a
self-autonomous system.
In fact, by introducing the variables $x$ and $y$ defined by $x={\sigma}'$ and
$y=\zeta^{-1}{\cal H}_*$,  we derive
\begin{eqnarray}
x^{\prime}&=&{3\zeta\over 4}
\left[
2\gamma x^2-4xy
+4\zeta^2 (4-3\gamma)y^2+\gamma
\right]
\label{SA1}
\\
y^{\prime}
&=&
{1\over 8\zeta}
\left[
-2(2-\gamma)x^2
+16\zeta^2 xy
-12\zeta^2 \gamma y^2
+\gamma
\right]
\,.
\nonumber \\
&&
\label{SA2}
\end{eqnarray}

By choosing $x'=y'=0$, we find
 four fixed-points in this system;
\begin{widetext}
\begin{eqnarray}
&&
{\rm FP1}_\pm: (x_F,y_F)=(x_1^{(\pm)},y_1^{(\pm)})
\equiv
\pm \left({2\zeta\over \sqrt{3-8\zeta^2}}, 
{1\over2\zeta  \sqrt{3-8\zeta^2}}\right)
\label{FP1}
\,,
\\
&&
{\rm FP2}_\pm: (x_F,y_F)=(x_2^{(\pm)},y_2^{(\pm)})
\equiv\pm \left({\sqrt{\gamma}\over 
\sqrt{2(2-\gamma-2(4-3\gamma)\zeta^2)}}, 
{\sqrt{\gamma}\over 
\sqrt{2(2-\gamma-2(4-3\gamma)\zeta^2)}}
\right)
\label{FP2}
\,.
\end{eqnarray}
\end{widetext}

\begin{figure}[ht]
\begin{center}
\mbox{}\\[-3.em]
\includegraphics[scale=.3]{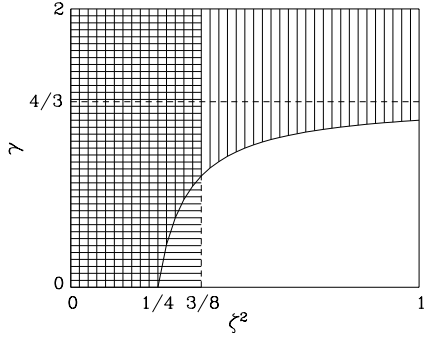}
\caption{The fixed points ${\rm FP1}_\pm$ and  ${\rm FP2}_\pm$ exist in
 the horizontally- and vertically-shaded  regions, respectively.  
}
\label{fixed_point}
\end{center}
\end{figure}

The fixed points ${\rm FP1}_\pm$ exist if $\zeta^2<3/8$,
while the fixed points ${\rm FP2}_\pm$ exist 
 if $\gamma \geq 4/3$ or if $\gamma < 4/3$ with 
$\zeta^2<{(2-\gamma)/[ 2(4-3\gamma)]}$
(equivalently, $\gamma>2(4\zeta^2-1)/(6\zeta^2-1)$).
For $\zeta=1/2$, two types of fixed points coincide to each other.
 In Fig. \ref{fixed_point}, we show in which portion of the
 $\zeta^2$-$\gamma$ plane we find  the fixed points.

From (\ref{Hamiltonian1}) with $k=0$, we obtain
\begin{eqnarray}
{\rho_*\over V_*}= 2(6\zeta^2 y^2-x^2)-1   
\label{enery_density}
\,,
\end{eqnarray}
which is constant at the fixed points. 

At the fixed point FP1$_\pm$, we find immediately
\begin{eqnarray}
(\rho_*/V_*)_{\rm FP1}=0, 
\end{eqnarray}
while at the fixed point FP2$_+$,
we obtain 
\begin{eqnarray}
\left({\rho_*\over V_*}\right)_{\rm FP2}= {2(4\zeta^2-1) \over 2-\gamma
-2(4-3\gamma)\zeta^2}.
\label{enery_density2}
\end{eqnarray}
This result is consistent with Eq. (\ref{em_cons2}) , i.e. 
\begin{eqnarray}
\left({\rho_*\over V_*}\right)^\prime=-3
\gamma\zeta(y-x)\left({\rho_*\over V_*}\right)
\,,
\end{eqnarray}
the right-hand side of which vanishes at the fixed point FP2$_+$,
showing that the ratio of the energy density to the potential is constant.

Next we discuss the scale factor and scalar field  at the fixed points.
We have 
\begin{eqnarray}
{\sigma}^\prime=x_F~,~~(\ln a_*)^\prime=\zeta y_F
\,, \label{sigmat}
\end{eqnarray}
for fixed points.

For FP1$_+$, we have $(x_F,y_F)=(x_1^{(+)},y_1^{(+)})$,  giving
\begin{eqnarray}
\sigma&=&{2\zeta\over \sqrt{3-8\zeta^2}}\tau_*+\sigma_0\,,
\\
a_*&=&a_{*0} \exp\left[{\tau_*\over 2\sqrt{3-8\zeta^2}}\right]
\,. \label{sigmat2}
\end{eqnarray}
By inverting (\ref{tau1}) and substituting from (\ref{sigmat}) we obtain
\begin{eqnarray}
t_*=t_{*0} \exp\left[{4\zeta^2\over \sqrt{3-8\zeta^2}}\tau_*
\right]
\,.\label{scaleft}
\end{eqnarray}
Substituting this back into (\ref{sigmat2}) and (\ref{scaleft}) then yields
\begin{eqnarray}
\sigma&=&{1\over 2\zeta}\ln \left({t_*\over t_{*0}}\right)
+\sigma_0
\,, \label{sigmabar}
\\
a_*&=&a_0 \left({t_*\over t_{*0}}\right)^{1/ (8\zeta^2)}
\,. \label{azero}
\end{eqnarray}
In order to fix ${\sigma}_0$, we  go back to the 
original equations of motion,
i.e,
setting $\rho_*=0$ finding
\begin{eqnarray}
H_*^2={1\over 3}\left(
{1\over 2}\dot{\sigma}^2+V_*
\right)
\,,
\end{eqnarray}
which gives  the value of the scalar field 
${\sigma}_0$ at $t_*=t_{*0}$ as
\begin{eqnarray}
\exp\left(-4\zeta{\sigma}_0\right)={3-8\zeta^2\over 64\zeta^4 t_0^2 V_0}
\,.
\end{eqnarray}
Eq. (\ref{azero}) shows that the solution with 
$\zeta<1/(2\sqrt{2})$ gives a power-law inflation.

For the fixed point FP2$_+$, we have
$(x_F,y_F)=(x_2^{(+)},y_2^{(+)})$, replacing (\ref{sigmabar}) and
(\ref{azero}) by
\begin{eqnarray}
\sigma&\hspace{-.3em}=\hspace{-.3em}&{\sqrt{\gamma}\over \sqrt{
2(2-\gamma-2(4-3\gamma)\zeta^2)}}\tau_*+{\sigma}_0
\,,
\\
a_*&\hspace{-.3em}=\hspace{-.3em}&a_{*0} \exp\left[{{\zeta \sqrt{\gamma}\over 
\sqrt{
2(2-\gamma-2(4-3\gamma)\zeta^2)}}\tau_*}
\right],
\end{eqnarray}
respectively.  The cosmic time in the Einstein 
frame is
\begin{eqnarray}
t_*=t_{*0} \exp\left[{{2\zeta \sqrt{\gamma}\over
\sqrt{2(2-\gamma-2(4-3\gamma)\zeta^2)}}\tau_*}
\right]
\,.
\end{eqnarray}
Hence  we find
\begin{eqnarray}
\sigma&=&{1\over 2\zeta}\ln \left({t_*\over t_{*0}}\right)
+{\sigma}_0
\,,\label{sigmardsol}
\\
a_*&=&a_{*0} \left({t_*\over t_{*0}}\right)^{1/2}
\,.
\label{scalerdsol}
\end{eqnarray}
It is important to notice that $a_* \sim t_*^{1/2}$ follows also for
dust-dominance.   In fact this behavior is true for any equation of state.
 For the remedy of this unfavorable result, the reader
is advised to see Section 4.4.3 of \cite{Fujii_Maeda} or Section 3.4 of 
\cite{ptpinv}.

To fix ${\sigma}_0$, we use the original Friedmann equation:
\begin{eqnarray}
H_*^2={1\over 6}\dot{\sigma}^2
+{V_*\over 3}
\left(
1+{\rho_*\over V_*}\right)
\,,
\end{eqnarray}
where ${\rho_*/V_*}$ is a constant given by Eq. (\ref{enery_density2}).
We then find
\begin{eqnarray}
\exp\left(-4\zeta{\sigma}_0\right)={(4\zeta^2-1)
[2-\gamma-2(4-3\gamma)\zeta^2]
\over 16\gamma\zeta^4 t_{*0}^2 V_0}
\,.
\label{sigmabar1}
\end{eqnarray} 
Note that spacetime in this range is static in the Jordan frame, as will
be shown in (\ref{aJ}).

\subsection{Stability Analysis 
and the Attractors}

Next we analyze stability of the fixed points FP1$_\pm$ and FP2$_\pm$ 
in the self-autonomous system (\ref{SA1}) and (\ref{SA2}).

\subsubsection{perturbation analysis}

The simplest way is to apply a perturbation analysis.
We perturb the variable $(x, y)$ around the fixed point $(x_F,y_F)$
as
\begin{eqnarray}
\left(
\begin{array}{c}
x\\
y
\end{array}
\right)=
\left(
\begin{array}{c}
x_F+\delta x\\
y_F+\delta y
\end{array}
\right)
\label{pertubation}
\,.
\end{eqnarray}  
Inserting Eq. (\ref{pertubation})
into the basic equations (\ref{SA1}) and (\ref{SA2}),
we find a set of the linear differential equations:
\begin{eqnarray}
\left(
\begin{array}{c}
\delta x\\
\delta y
\end{array}
\right)^\prime=
\left(
\begin{array}{cc}
A_{xx} & A_{xy}\\
A_{yx} & A_{yy}
\end{array}
\right)\left(
\begin{array}{c}
\delta x\\
\delta y
\end{array}
\right)
\,,
\label{perturbation_eq}
\end{eqnarray} 
where
the components of the matrix $A$ is given by
\begin{eqnarray}
\left\{
\begin{array}{l}
A_{xx}=
3\zeta(\gamma x_F- y_F)
\\[1em]
A_{xy}=3\zeta(-x_F+2\zeta^2(4-3\gamma)y_F)\\[1em]
A_{yx}=-{\mbox{\small $2-\gamma$}\over \mbox{\small $2\zeta$}}x_F+
2 \zeta y_F\\[1em]
A_{yy}=\zeta(2 x_F-3\gamma y_F)
~~.
\end{array}
\right.
\end{eqnarray}
Assuming 
$
\delta x, 
\delta y
\propto e^{\omega \tau_*}
$, 
we find the equation for the eigenvalue $\omega$ as
\begin{eqnarray}
\omega^2 - {\rm Tr} A ~\omega + \det A =0
\label{eigen_eq}
\,,
\end{eqnarray}
where 
\begin{eqnarray}
{\rm Tr} A &=&{1\over 4\zeta^2-1}\left[
(x_F-4\zeta^2 y_F)
\nonumber 
\right.\\
&+&
\left.
\{3\gamma(4\zeta^2-1)-(3-8\zeta^2)\}(x_F-y_F)
\right] \,,
\label{trace}
\\
 \det A &=&{3\over 2(4\zeta^2-1)}
\left[\left(2-\gamma-2(4-3\gamma)\zeta^2\right)(x_F-4\zeta^2 y_F)^2
\nonumber 
\right.\\
&-&
\left.
2\gamma\zeta^2(3-8\zeta^2)(x_F-y_F)^2\right] \,.
\label{det}
\end{eqnarray}
In this expression, the first terms in (\ref{trace}) and (\ref{det}) vanish 
for FP1$_\pm$, while the second terms disappear for FP2$_\pm$.  The
fixed point is stable in the following  two cases:
\begin{itemize}
\item Eq. (\ref{eigen_eq}) has two negative real roots.
\item Eq. (\ref{eigen_eq}) has 
two complex conjugate roots with a negative real part.
\end{itemize} 
The condition is 
\begin{eqnarray}
{\rm Tr} A <0,~~{\rm and}~~~{\rm det} A >0 
\label{stability_condition}
\,.
\end{eqnarray}
For FP1$_\pm$, we find 
\begin{eqnarray}
({\rm Tr} A)_{{\rm FP1}_\pm} 
&=&
[4(3\gamma+2)\zeta^2-3(\gamma+1)]\,y_1^{(\pm)}
\label{trace1}
\\
 (\det A)_{{\rm FP1}_\pm} &=&3
\gamma\zeta^2(3-8\zeta^2)(4\zeta^2-1)(y_1^{(\pm)})^2
\label{det1}
~~~~
\end{eqnarray}
For the expanding universe (FP1$_+$), which we are interested in,
the above condition (\ref{stability_condition}) gives
\begin{eqnarray}
\zeta^2<{1\over 4}
\,.
\end{eqnarray}

For FP2$_\pm$, we have 
\begin{eqnarray}
({\rm Tr} A)_{{\rm FP2}_\pm} &=&-y_2^{(\pm)}
\label{trace2}
\\
 (\det A)_{{\rm FP2}_\pm} &=&
-3\gamma\zeta^2(3-8\zeta^2)(4\zeta^2-1)(y_2^{(\pm)})^2
\label{det2}
~~~~~~~
\end{eqnarray}
The stability condition for the expanding universe (the fixed point
FP2$_+$) gives
\begin{eqnarray}
\zeta^2>{1\over 4}
\,.
\end{eqnarray}
\begin{figure}[ht]
\begin{center}
\mbox{}\\[-2.0em]
\includegraphics[scale=.3]{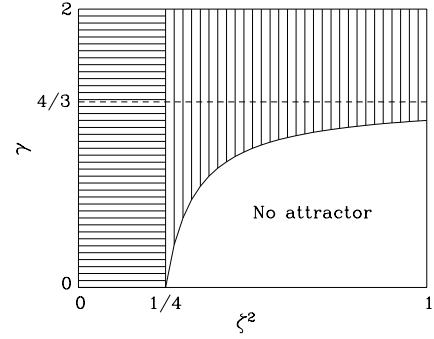}
\caption{The attractor fixed-points 
 (FP1$_+$ and FP2$_+$) in the same parameter space of $\zeta^2$ and
 $\gamma$, also the same shading pattern as in Fig. \ref{fixed_point}.}
\label{attractor}
\end{center}
\end{figure}
In this way we find that FP1$_+$ and FP2$_+$ are the attractors 
for $\zeta^2<{1/ 4}$ and for $\zeta^2>{1/ 4}$
(also with $\zeta^2<(2-\gamma)/[2(4-3\gamma)]$
for the existence of FP2$_+$),
respectively.
For $\zeta^2=1/4$, two types of fixed points merge with each other, sharing the same
behaviors, but like a saddle point rather than an attractor.
Fig. \ref{attractor} shows in which portion of the $\zeta^2$-$\gamma$ plane 
we have the attractor fixed points.

\subsubsection{phase space analysis}

We also study stability by use of a phase-space analysis of the dynamical
system with (\ref{SA1}) and (\ref{SA2}) 
\cite{phase_space_analysis1,phase_space_analysis2,phase_space_analysis3,phase_space_analysis4}.  
We may discuss global
stability  rather than local one in the perturbative approach, 
as we will show shortly.  There is a limitation, however, because
dependence on the values of $\gamma, \zeta^2$ is not as simple as shown in
(\ref{trace1}) and in the subsequent equations.  We must develop the
computation for each of these parameters separately, though without any
difficulty in principle.  For this reason, we show the following examples of
radiation-dominance ($\gamma =4/3$) illustrating generic features shared
commonly by this type of analyses with any values of the parameters.

For the sake of convenience we start with reproducing (\ref{SA1}) and  (\ref{SA2})
for $\gamma =4/3$;
\begin{eqnarray}
&&
x'=\zeta \left( 2x^2 -3xy +1 \right),
\label{rad3_5}\\
&& 
y'=\frac{1}{2\zeta}\left( -\frac{1}{3}x^2 +4\zeta^2  xy -4\zeta^2 y^2 
+\frac{1}{3}\right),
\label{rad3_6}
\end{eqnarray}
which determine how the point ($x,y$) representing the solution moves with
time.  The fixed points are (\ref{FP1}) for  FP1$_\pm$, while  
$
(x_2^{(\pm)},
y_2^{(\pm)})=\pm (1,1)
$
from (\ref{FP2}) for FP2$_\pm$.
We also focus upon FP2$_\pm$.

A basis for the required  analysis is prepared first by drawing the ``null
 curves'' for $x'=0$ and $y'=0$.  By choosing the vanishing left-hand
 sides of (\ref{rad3_5}) and (\ref{rad3_6}), we find that the former
 curves, solid (blue), are 
 in fact hyperboloids, whereas the latter ones, dashed (red) curves are
 either hyperbolic or elliptic depending on $\zeta^2>1/3$ or
 $\zeta^2<1/3$, 
respectively, as shown in Figs. \ref{f1} and \ref{f2}, used
 separately for the analyses of the two choices for $\zeta^2$.  
The crossings represent fixed points.
Note that there are four fixed points (FP1$_\pm$, FP2$_\pm$) for
 $\zeta^2<1/3$, while two fixed points (FP2$_\pm$) for  $\zeta^2>1/3$.

\begin{widetext}
\mbox{}\\[-8.5em]
\hspace*{-1.5em}
\begin{minipage}{6.cm}
\hspace*{-1.0em}
\includegraphics[keepaspectratio,width=9.4cm]{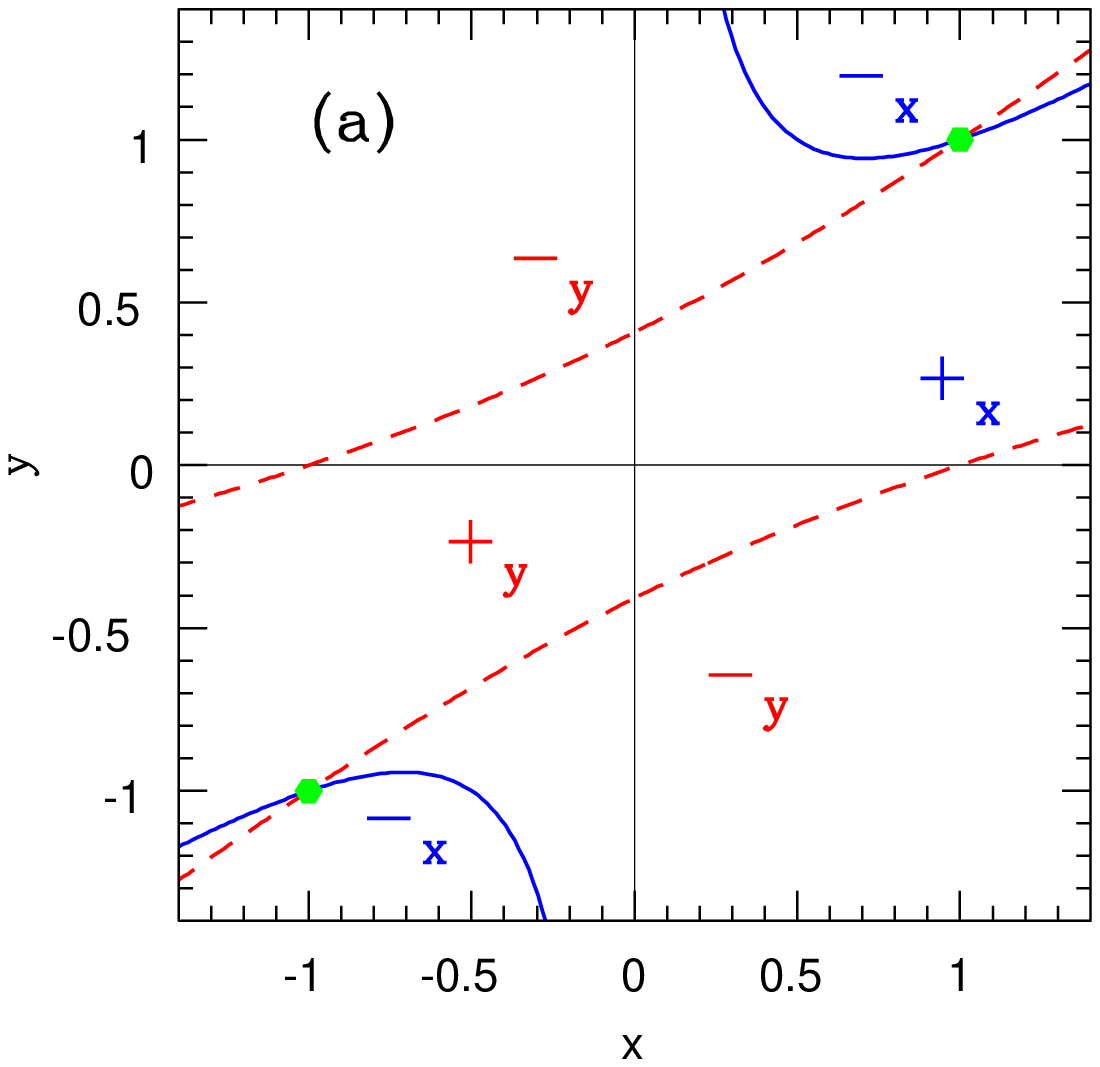}
\end{minipage}
\hspace{-.6em}
\begin{minipage}{6.cm}
\hspace*{-1.0em}
\includegraphics[keepaspectratio,width=9.4cm]{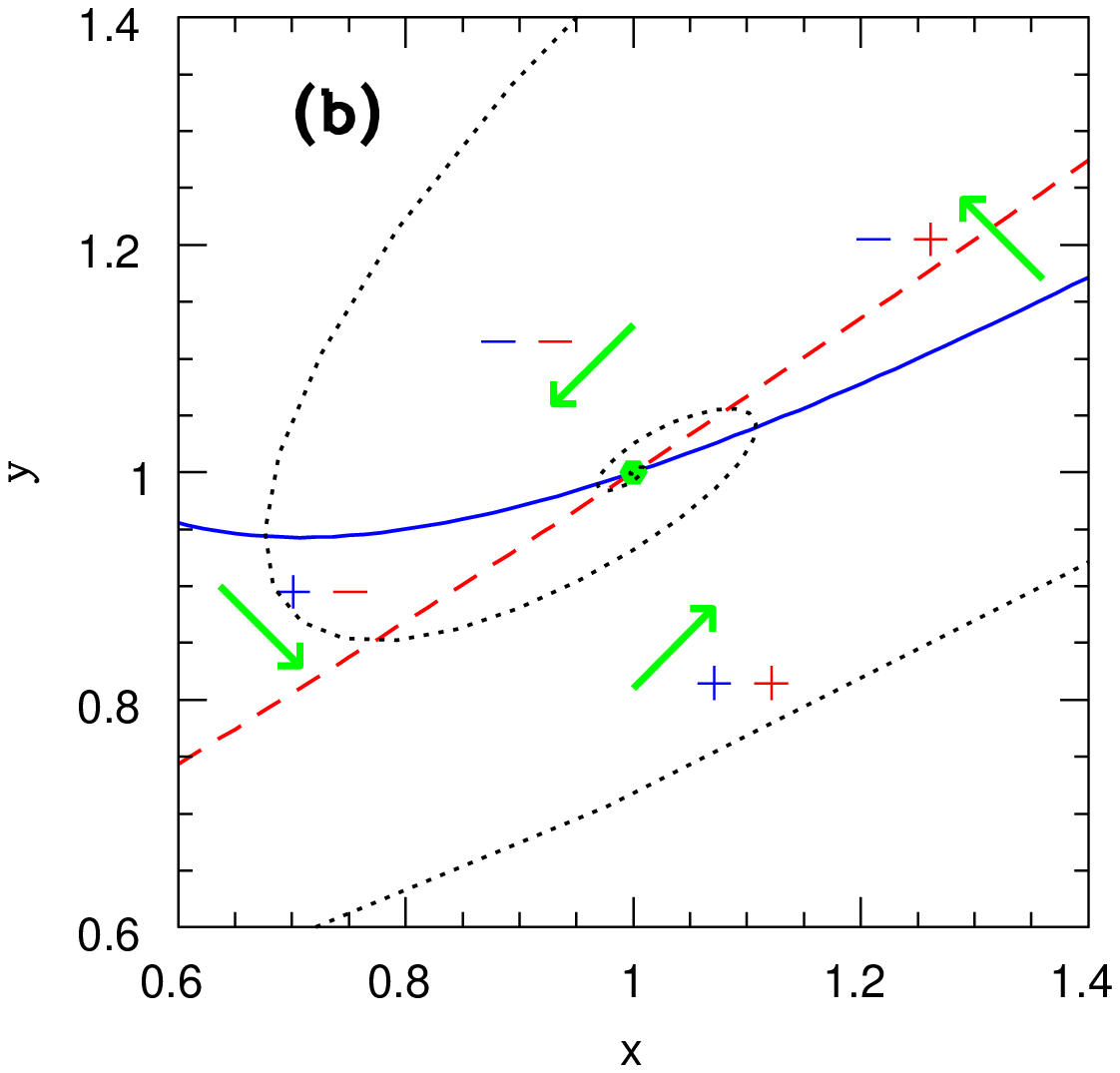}
\end{minipage}
\hspace{-.6em}
\begin{minipage}{6.cm}
\hspace*{-1.0em}
\includegraphics[keepaspectratio,width=9.4cm]{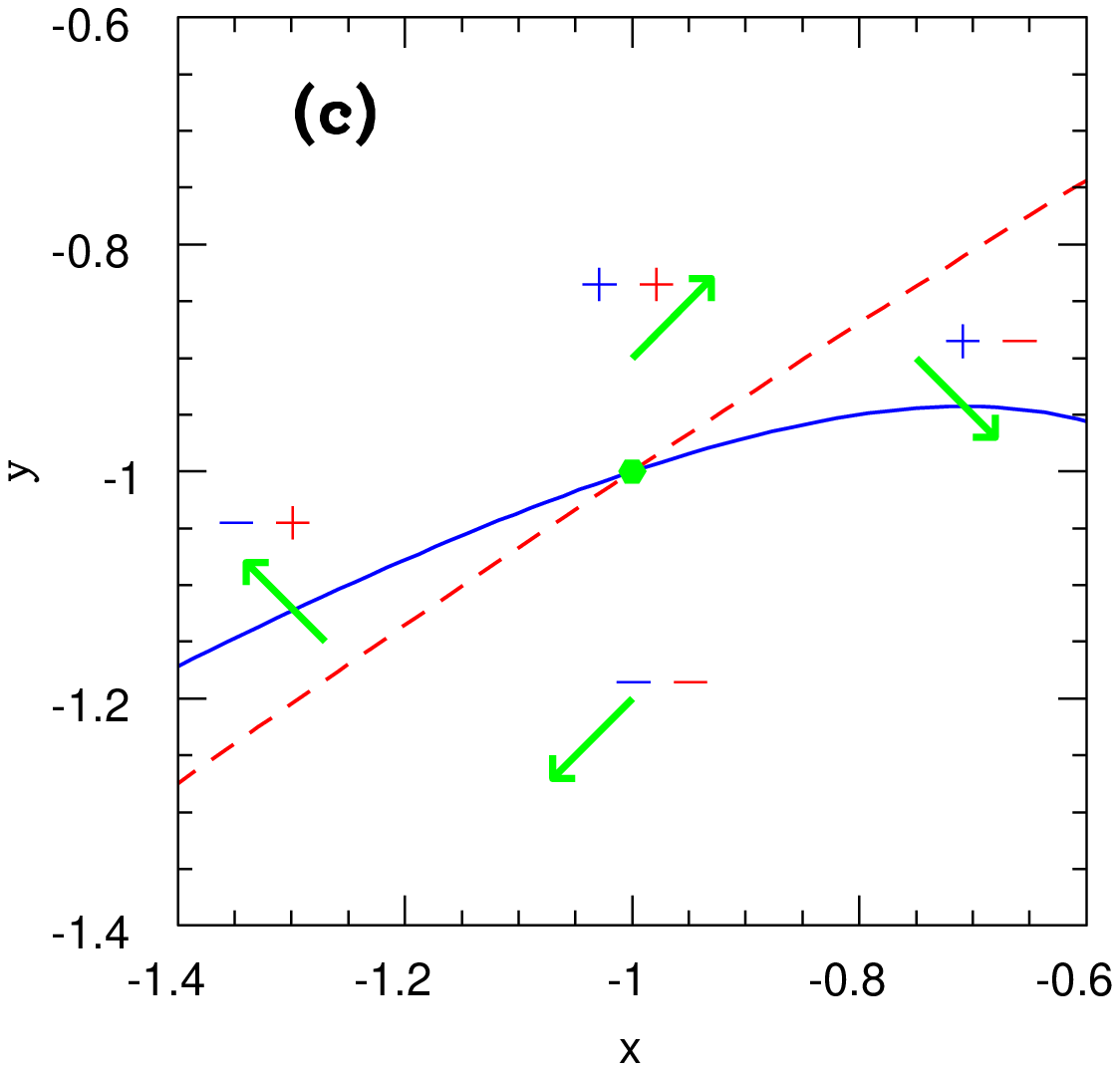}
\end{minipage}
\mbox{}\\[-3.0em]
\begin{figure}[ht]
\caption{\small An example of $\zeta^2>1/3$ in the phase diagram of
 $x={\sigma}'$ and $y=\zeta^{-1}{\cal H}_*$.  
In the overall diagram
 (a), we chose an example of  $\zeta^2 =1/2$ to draw solid (blue) and
 dashed (red) curves for the null curves for $x'=0$ and $y'=0$,
 respectively.  They separate regions of different signs of them,
 denoted by $+_{\rm x}$ and $+_{\rm y}$, 
for $x'>0$ and $y'>0$, respectively, for example.  The crossings, or
 fixed points, occur between two hyperboloids, marked with a blob at
FP2$_+$($x=y=1$) as an attractor, and the one at FP2$_-$($x=y=-1$)
 as a repeller (no fixed points FP1$_\pm$ for this parameter).
  In each of smaller regions bounded by the null curves, as shown in (b) and
 (c), close-up views near the attractor and the repeller, respectively,
 we have directions  of ``flows'' of solutions, or trajectories, shown
 symbolically by  the arrows together  with nearby pair of signs for the
 increment ($+$) or  decrement ($-$) of the $x$-  and $y$-components denoted in this order.  A
 dotted curve in (b) is a  trajectory  starting from the initial values
 $\ln \rho =0.1, \varphi =0.1,  \dot{\varphi} =0$ at the initial Jordan  frame ``time'' $\ln t=1$,  where $\varphi =(\xi /2)\phi^2$ with $\xi=1/4$.
  The trajectory  flows basically in  accordance with the arrows,
 entering the frame of diagram near the lower-left corner, traversing to
 go outside beyond the right edge, re-entering at the top, spiraling finally into the attractor.} 
\label{f1}
\end{figure}
\end{widetext}

In Fig. \ref{f1} (a), we draw two sets of hyperboloids for
 $\zeta^2 > 1/3$.
 It is rather easy to determine which side of each curve implies
 positive or negative $x'$ and $y'$, as shown by the symbols
 like $+_{\rm x}$ and $+_{\rm y}$, 
for $x'>0$ and $y'>0$, respectively.  We then determine in
 what direction a point, or better called a trajectory, ``flows'' with
 time in a given region in  $x$-$y$ space bounded by the null
 curves, as illustrated  symbolically by arrows (green). 
In Fig. \ref{f1} (b), we show a typical trajectory represented by a
dotted curve which enters the diagram first near the lower-left corner,
making a big loop outside the frame of the diagram, then re-entering
again, finally  spirals into the crossing at FP2$_+$ ($x=y=1$), which   
 corresponds to (\ref{sigmardsol})  and  (\ref{scalerdsol}).
This is the way  we now 
 establish our previous numerical results obtained on the basis of the
 heuristic approach \cite{Fujii_Maeda},\cite{ptpinv} to be an authentic
 attractor in a strict sense.

We notice, on the other hand, that a trajectory, or the solution, may
not always converge to a fixed point, straying instead toward infinity,
as will be illustrated in Appendix B.  Obviously, however, not reaching
the point of $x'=y'=0$ implies the destination not corresponding to the
steady and lasting solution, as given by (\ref{sigmardsol}) and
(\ref{scalerdsol}), for example.  In other words, any solution that
survives a long time must come from the attractor.

There is another fixed point  FP2$_-$ 
($x=y=-1$) to which no trajectory flows into
 as long as we start with a positive $\rho$, the same sign as
 $V_0$.  Note that this fixed point corresponds to the contracting
 universe.  The  flows shown in  Fig. \ref{f1} (c)
 indicates that this is a
 repeller to be interpreted as the time-reversed  point against the
 attractor at  FP2$_+$.

We encounter another  complication, on 
 the other hand, if $\zeta^2<1/3$, for which 
 we have four crossings as illustrated in Fig. \ref{f2}(a).  Focusing upon
 the behaviors in the upper-right quadrant, we have magnified views at
 each of the two, one in (b) around $x=y=1$, corresponding to the
 fixed-point solution of FP2 given by (\ref{sigmardsol})-(\ref{scalerdsol})
 above, and another shown in (c) now categorized into the type of FP1. 
 The flows in (b) do indicate that the
 the solution in all directions tends finally to the crossing, a real
 attractor, while according to those illustrated in (c) there are some
  narrow strips sandwiched between two null curves in which the
 trajectory tends away from the crossing.  This reminds us of a
 saddle-point potential rather than the purely attractive or repulsive
 potential in mechanics, hence implying an unstable solution which may
 not survive a long time, leading ultimately to the same drift toward
 infinity, as was remarked toward the end of the discussion of the
 preceding example in Fig. \ref{f1}.   For $1/4<\zeta^2<1/3$, therefore, we may fail
 to reach the attractor  solution with some probability depending on the
 initial values, or on what portion in the $x$-$y$ plane we started off.  We
 may constrain ourselves to $\zeta^2 >1/3$, though another {\it a
 posteriori} attitude might be suggested:  Given what we  are at
 present,   right  initial values, or right initial locations in phase
 space,  must have been chosen  to reach the attractor solution 
 (\ref{sigmardsol})-(\ref{scalerdsol}). 

\begin{widetext}
\mbox{}\\[-9.5em]
\hspace*{-1.5em}
\begin{minipage}{6.cm}
\hspace*{-1.0em}
\includegraphics[keepaspectratio,width=9.4cm]{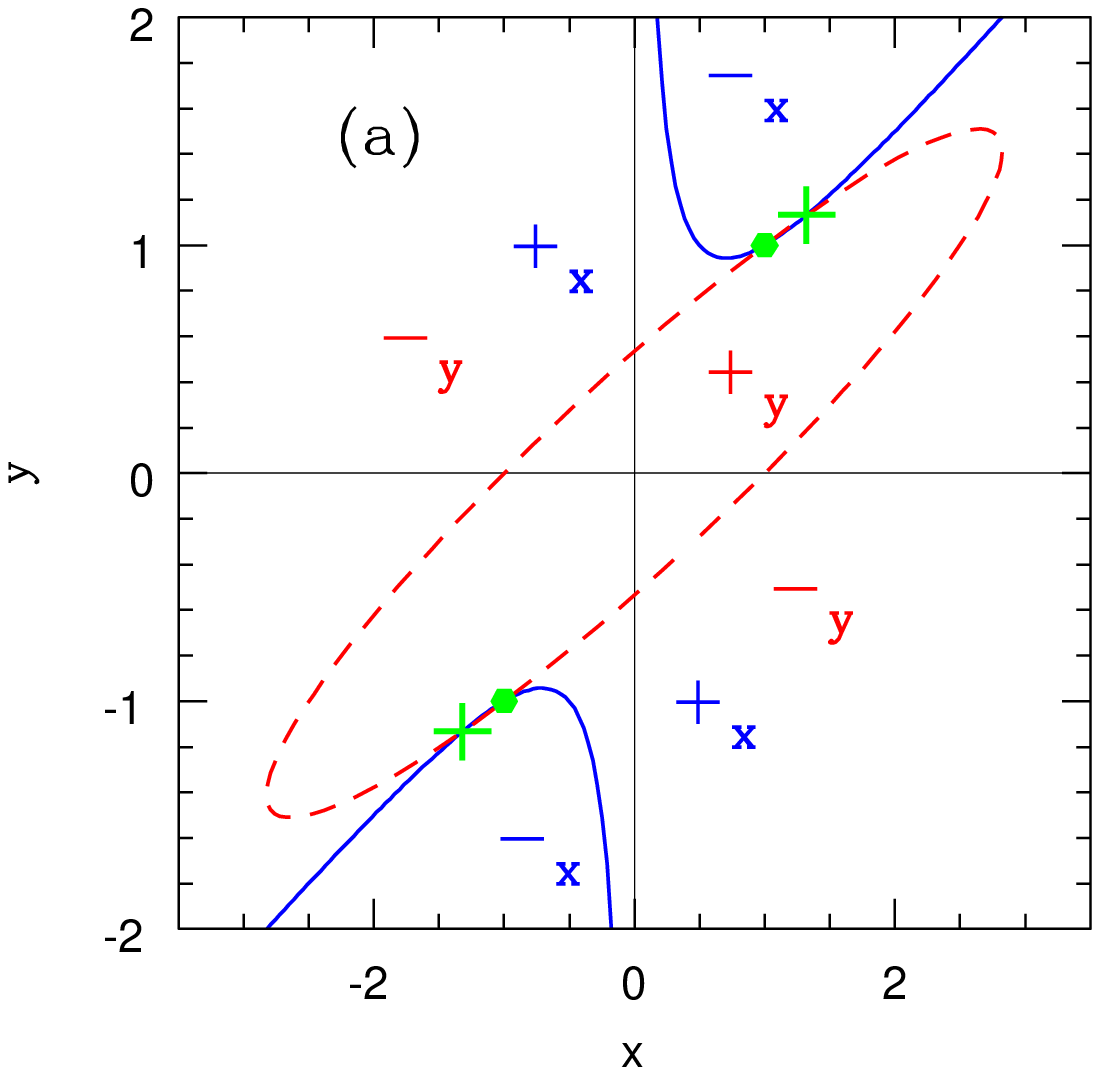}
\end{minipage}
\hspace{-.6em}
\begin{minipage}{6.cm}
\hspace*{-1.0em}
\includegraphics[keepaspectratio,width=9.4cm]{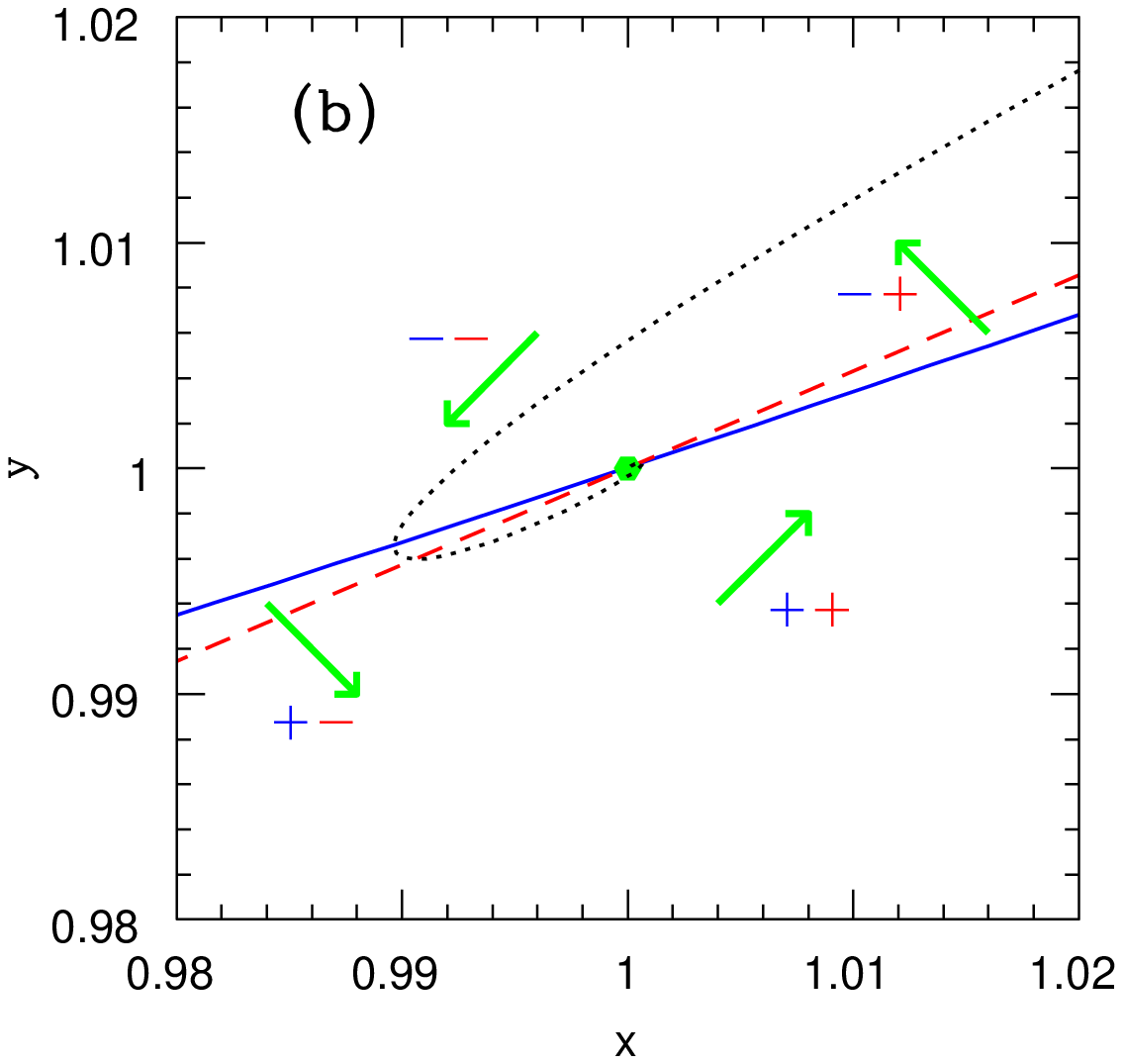}
\end{minipage}
\hspace{-.6em}
\begin{minipage}{6.cm}
\hspace*{-1.0em}
\includegraphics[keepaspectratio,width=9.4cm]{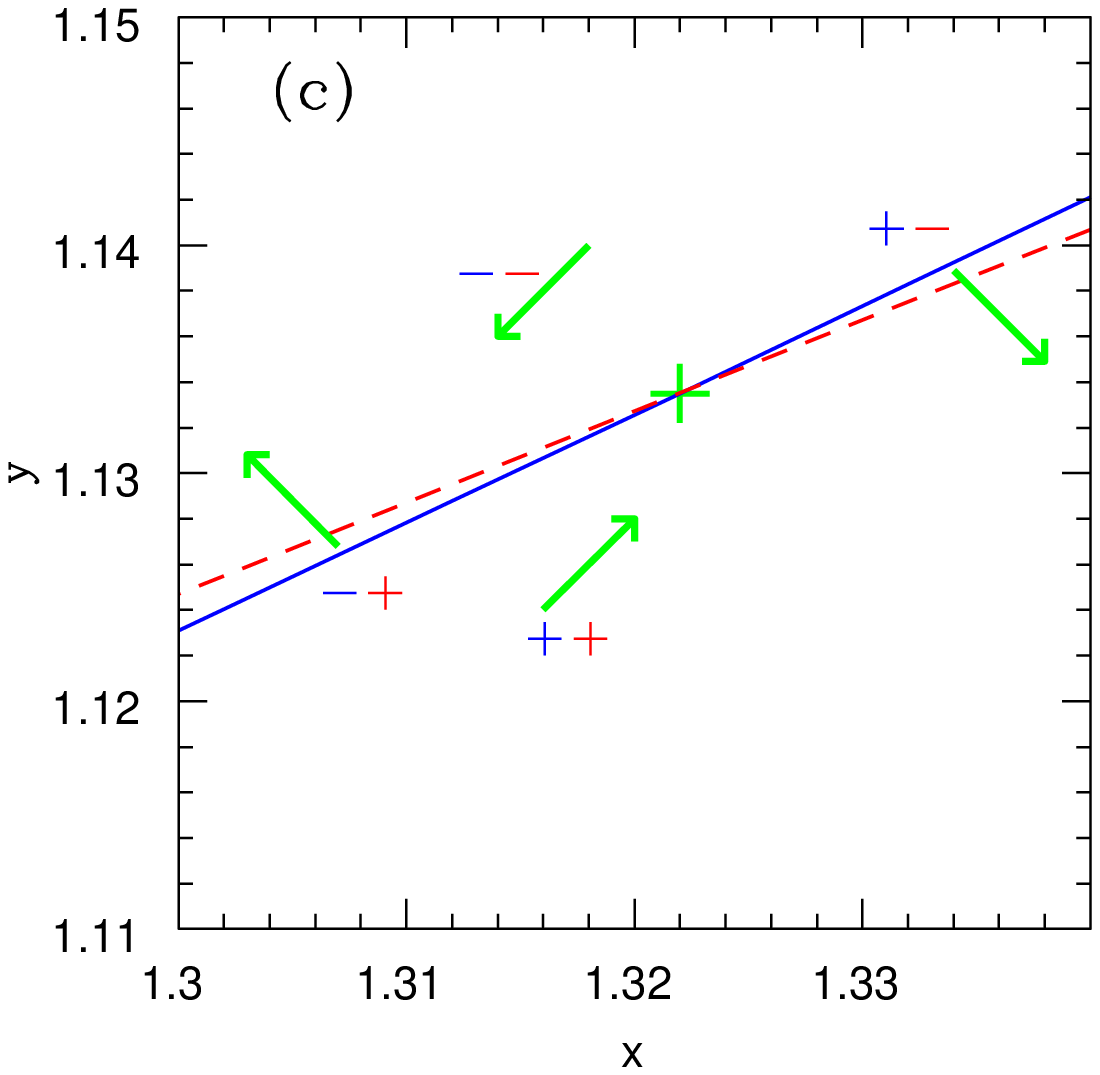}
\end{minipage}
\mbox{}\\[-2.0em]
\begin{figure}[h]
\caption{An example of $1/4<\zeta^2<1/3$. In the overall diagram (a), we
 chose an example of $\zeta^2=0.2916, \xi=0.3890$.  Unlike in Fig. \ref{f1},
 the dashed (red) null curve is an ellipsoid, thus producing four
 crossings.  In addition to the attractor at  FP2$_+$($x=y=1$),
 denoted by a
 blob (green), we have another at FP1$_+$ ($x=1.322, y=1.134$)
 marked by a cross
 (green), both accompanied with the time-reversal counterparts in the
 left-lower quadrant (FP1$_-$ and FP2$_-$), 
which we ignore for brevity.  In a close-up view
 (b) around FP2$_+$, we show an example of a trajectory with the same
 initial values as considered in Fig. \ref{f1} (b),  certainly spiraling into
 the fixed point in accordance with the arrows of flows, while we find no
 trajectories in (c) around another crossing of the type FP1$_+$.  The
 behaviors of the arrows in the right-upper and left-lower strips  remind us 
of a saddle-point potential resulting in unstable motion.  }
\label{f2}
\end{figure}
\end{widetext}

\begin{figure}[ht]
\mbox{}\\[-6.6em]
\hspace*{2.0em}
\includegraphics[scale=.42]{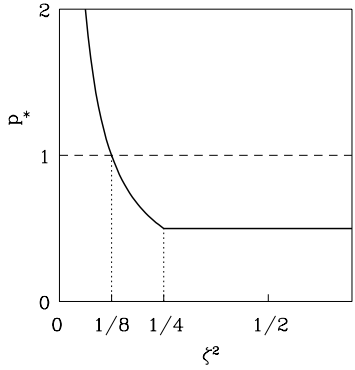}
\mbox{}\\[-2.4em]
\caption{The power index $p_*$ of the scale factor in the Einstein frame
for the attractor solutions.
The power-law inflation appears when $\zeta^2<1/8$.
The power exponent is always 1/2 for FP2$_+$, 
independent of $\gamma$.
}
\label{power}
\end{figure}

We apply the same analysis for any values of $\gamma$, 
finding the similar results.
We summarize the power index  of scale factor  
of the the attractor solutions
(FP1$_+$ for $\zeta<1/2$, and FP2$_+$ for $\zeta>1/2$)
in Fig. \ref{power}.
We find that the power-law inflation appears when $\zeta^2<1/8$.
The power exponent of the scale factor in the Einstein frame is
always 1/2 for FP2$_+$, which does not depend on the equation of state
of the matter fluid, as derived in (\ref{scalerdsol}).
\section{cosmology in  the Jordan frame}

We repeat the similar analysis for  the same system as in Section III
 now in the Jordan frame.

From (\ref{action_Jordan}) together with the FRW metric
\begin{eqnarray}
ds^2=-dt^2+a^2 ds_3^2
\,,
\label{lineJ}
\end{eqnarray}
we derive
\begin{eqnarray}
&&
3\xi\phi^2 \left(H^2+{k\over a^2}\right)+6\xi H
\phi \dot{\phi}
=
{\epsilon\over 2}\dot{\phi}^2+V_0+\rho 
~~~~
\label{Friedmann_ST}
\\
&&
\xi \left(\phi \ddot{\phi}+ \dot{\phi}^2
+3H\phi \dot{\phi} \right)
=
\zeta^2\left(4V_0+\rho-3P\right)
\label{scalar_ST}
\\
&&
\dot{\rho}+3H(P+\rho)=0
\label{matter_ST}
\end{eqnarray}
where 
\begin{eqnarray}
H=\frac{\dot{a}}{a}
\,,
\end{eqnarray}
while $P=P_* \exp (4\zeta\sigma)$ and $\rho=\rho_*\exp (4\zeta\sigma)$  are the pressure and the
energy density, respectively, in the Jordan frame.  Note that the quantities in the Jordan
frame are all denoted unmarked in contrast to those marked with the
subscript $*$ in the Einstein frame.  Accordingly, the dots in
(\ref{Friedmann_ST})-(\ref{matter_ST}) imply differentiation with
respect to $t$ instead of $t_*$.

Since $ds^2=\Omega^{-2}ds_*^2$ with $\Omega^2=\xi \phi^2
=\exp (2\zeta\sigma)$, we have the relations between the variables in Jordan frame and 
those in the Einstein frame as
\begin{eqnarray}
&&
dt=\Omega^{-1}dt_*\,,
\label{change_time_coordinate}\\
&&
a=\Omega^{-1} a_*
\label{change_scale_factor}\,.
\end{eqnarray}
Note that we do not change the coordinate system when we perform
a conformal transformation. However, since we use a cosmic time in each 
frame, we have to change the time coordinate between $t_*$ and $t$
according to (\ref{change_time_coordinate}). 
The Hubble expansion parameters $H$ and $H_*$ are defined 
by each cosmic time as
\begin{eqnarray}
H={1\over a}{d  a\over  dt}~,~~H_*={1\over a_*}{d a_* \over dt_*}
\,.
\end{eqnarray}
Hence we have the relation 
\begin{eqnarray}
H_*=\Omega^{-1}\left(H+{d\ln \Omega\over dt}
\right)
\,.
\end{eqnarray}

Introducing new time coordinate $\tau$
and and new scalar field $\Phi$, which are defined by
\begin{eqnarray}
d\tau&=&2 {\xi}^{-1/2} V_0^{1/2}{\phi}^{-1}\,dt
\\
{\Phi}&=&\ln \phi
\,,
\end{eqnarray}
respectively, we rewrite
Eqs. (\ref{Friedmann_ST})-(\ref{matter_ST})  as
\begin{eqnarray}
&&6 {\cal H}^2
+{3k\xi e^{2\Phi} \over 2  V_0 a^2}
=
{1\over \zeta^2}(\Phi^\prime)^2+{1\over 2}\left(
1+{\rho\over V_0} \right)
\label{Friedmann_ST2}
\\
&&
{\Phi}^{\prime\prime}
-2({\Phi}^\prime)^2+3{\cal H}{\Phi}^\prime=\zeta^2\left[1+
{(4-3\gamma)\rho\over 4V_0}
\right]~~~
\label{scalar_ST2}
\\
&&
\rho^\prime+3\gamma ({\cal H}-{\Phi}^\prime)
 \rho=0
\label{matter_ST2}
\,,
\end{eqnarray}
where 
\begin{eqnarray}
{\cal H}\equiv 
{{a}^\prime \over a}
+{\Phi}^\prime
\,,
\end{eqnarray}
and
the prime is the derivative with respect to $\tau$. 
We also assume the equation of state, $P=(\gamma-1)\rho$.

Now, we consider only the spatially flat universe, i.e, 
$k=0$ (See Appendix A for $k\neq 0$).
Taking the derivative of Eq. (\ref{Friedmann_ST2})
and using Eqs. 
(\ref{Friedmann_ST2}), (\ref{scalar_ST2}) and (\ref{matter_ST2}), we find 
\begin{eqnarray}
&&{\cal H}^\prime={\gamma-2\over 4\zeta^2}(\Phi^\prime)^2
+2{\cal H}\Phi^\prime
-{3\gamma\over 2}{\cal H}^2
+{\gamma\over 8}
\,,
\label{Friedmann_ST4}
\\
&&
{\Phi}^{\prime\prime}=
{3\gamma\over 2}({\Phi}^\prime)^2-3{\cal H}{\Phi}^\prime+
3(4-3\gamma)\zeta^2{\cal H}^2+{3\gamma\over 4}\zeta^2
\,.
\nonumber \\
&&
\label{scalar_ST4}
\end{eqnarray}
This is again a self-autonomous system with two variables $x$ and $y$:
\begin{eqnarray}
&&
x'={3\zeta\over 4}\left[
2\gamma x^2-4xy+
4\zeta^2(4-3\gamma)y^2
+\gamma
\right]
\,,
\\
&& y'={1\over 8\zeta}\left[
-2(2-\gamma)x^2+
16\zeta^2 xy-12\gamma\zeta^2 y^2+ \gamma 
\right]
\,.
\nonumber \\
&&~
\end{eqnarray}
where $x=\zeta^{-1}\Phi^\prime$
and $y=\zeta^{-1}{\cal H}$.
These equations turn out to be precisely the same as Eqs. (\ref{SA1})
and (\ref{SA2}), respectively, implying  the same dynamical system, sharing 
 the same fixed points: FP1$_\pm$ (\ref{FP1}) and FP2$_\pm$(\ref{FP2}).

The energy density is given by
\begin{eqnarray}
{\rho\over V_0}&=&12{\cal H}^2-{2\over \zeta^2}(\Phi')^2-1
\nonumber \\
&=&12\zeta^2y^2-2x^2-1
\,,
\end{eqnarray}
precisely the same as (\ref{enery_density}) for the Einstein frame.

We only show the explicit solutions of the fixed points,
because the dynamical properties such as an attractor
is the same as that in the Einstein frame.
For FP1$_\pm$, we have 
\begin{eqnarray}
a&=&a_0 (\tau-\tau_0)^{{\mbox{\small $1$}\over \mbox{\small $4\zeta^2$}}-\mbox{\small $1$}}
\nonumber \\
 \phi&=& \pm {2\zeta\over \sqrt{3-8\zeta^2}}(\tau-\tau_0)
\\
\rho&=&0
\nonumber 
\,.
\end{eqnarray}
This gives a power-law  inflation for $\zeta^2<1/8$.

For FP2$_\pm$ ,
we have 
\begin{eqnarray}
a&=&a_0 \,,
\label{aJ} \\
\phi&=& \pm {\sqrt{\gamma}\over 
\sqrt{2(2-\gamma-2(4-3\gamma)\zeta^2)}}(\tau-\tau_0) \,,
\label{phiJ}
\\
\rho&=&{2(4\zeta^2-1)\over 2-\gamma-2(4-3\gamma)\zeta^2} \, V_0 \,,
\label{rhoJ}
\end{eqnarray}
where $a_0$ and $\tau_0$ are integration constants.
The spacetime is a static Minkowski space, in contrast with the
expansion in the Einstein frame as shown by (\ref{scalerdsol}).

The vacuum solution in radiation-dominance in \cite{dolgov} may
be interpreted as the limit $\zeta^2 \rightarrow 1/4$ in
(\ref{aJ})-(\ref{rhoJ}).  The constant $a$ was also derived in 
\cite{wett}.

\section{cosmology with a power-law potential}
The analysis in the preceding section can be readily extended to include
 the power-law potential,
\begin{eqnarray}
V(\phi)=\lambda_\alpha \,
 \phi^{\alpha}
\,.
\label{power-law potential}
\end{eqnarray}
in the Jordan frame.  An example is provided by extending  the action
(\ref{string}) in string theory to
\begin{eqnarray}
S={1\over 2}\int d^Dx \sqrt{-g} e^{-2\Phi}\left[R(g)-2\Lambda
+4\left(
\nabla\Phi\right)^2\right]
\,,
\end{eqnarray}
in which we have added the term of $\Lambda$, to be corresponded to
(\ref{power-law potential}) by choosing $\alpha=2$ and $\lambda_2=\Lambda/4$.

The action equivalent to (\ref{power-law potential}) but expressed in
the Einstein frame is given by
\begin{eqnarray}
S&=&\int d^4x \sqrt{-g_*}
\left[{1 \over 2} R_*(g_*)-{1\over 2}(
\nabla_*\sigma)^2-V_*(\sigma)
\right]
\nonumber \\
&+&\int d^4x \sqrt{-g_*}L(\sigma, \psi_*, g_*)
\,,
\end{eqnarray}
where
\begin{eqnarray}
V_*=\exp[{-\zeta(4-\alpha)\sigma}]\,V_0
\,,
\end{eqnarray}
with $V_0=\lambda_\alpha/|\xi|^{\alpha/2}$.
Since this is of the same  form as Eq. (\ref{Eaction}), we repeat the same
analysis as before.  One of the  points to be kept in mind is that the
conformal transformation is always the same as (\ref{cftmetric}) with
(\ref{Omega}).   Hence we have the same energy-momentum conservation law 
(\ref{energy_conservation_Einstein}), as well.

The basic equations for cosmology are then;
\begin{eqnarray}
&&
H_*+{k\over a_*^2}={1\over 3}\left(
\dot{\sigma}^2+V_*+\rho_*
\right),
\\
&&
\ddot{\sigma}+3H_*\dot{\sigma}
+{\partial V_*\over \partial \sigma}
=\zeta(\rho_*-3 P_*), 
\\
&&
\dot{\rho}_* +3H_*(P_*+\rho_*)=
- \zeta\dot{\sigma}(\rho_*-3P_*). 
~~~
\end{eqnarray}
Introducing a new dimension-free time coordinate as
\begin{eqnarray}
d\tau_*=2 \sqrt{{V}_*}\,dt_*,
\end{eqnarray}
also assuming the equation of state and focusing on $k=0$ as before, we 
obtain formally the
same equations as (\ref{Hamiltonian1})-(\ref{em_cons1}), but 
replace (\ref{em_cons2}) by
\begin{eqnarray}   
(\rho_*/V_*)'=-
\left[3\gamma {\cal H}_*-\zeta(3\gamma -\alpha) 
\sigma'\right](\rho_*/V_*)
\label{em_cons3}
\,.
\end{eqnarray}

\begin{widetext}
Computing in the same way as before,  we come to
 replacing (\ref{SA1}) and (\ref{SA2}) by
\begin{eqnarray}
&&
x'={\zeta\over 4}\left[
2\left(3\gamma-\alpha
\right)x^2
+12\zeta^2(4-3\gamma)y^2
-12 xy
+3\gamma-\alpha
\right],
\label{scalar_power3}
\\
&&
y'={1\over 8\zeta}\left[
-2(2-\gamma) x^2
-12\gamma\zeta^2 y^2+4\zeta^2\left(4-\alpha\right)xy+\gamma
\right]
\label{Friedmann_power3}
\,,
\end{eqnarray}
respectively, where $x={\sigma}'$ and $y=\zeta^{-1}{\cal H}_*$ as before.

We find four fixed points as the previous case:
\begin{eqnarray}
&&
{\rm FP1}_\pm: (x_F,y_F)=(x_1^{(\pm)},y_1^{(\pm)})
\equiv
\pm {1\over \sqrt{2(6-\zeta^2(4-\alpha)^2)}}
\left(\zeta(4-\alpha), 
{1\over\zeta }\right)
\label{FP_power1}
\,,
\\
&&
{\rm FP2}_\pm: (x_F,y_F)=(x_2^{(\pm)},y_2^{(\pm)})
\equiv\pm 
{1\over \sqrt{6[3\gamma(2-\gamma)-2(4-3\gamma)(3\gamma
-\alpha)\zeta^2
]}}
\left(
3\gamma,
3\gamma-{\alpha}
\right)
\label{FP_power2}
\,.
\end{eqnarray}
\end{widetext}

The fixed points FP1$_\pm$ exist if 
\begin{eqnarray}
4\left(
1-{\sqrt{6}\over 4\zeta}
\right)
<\alpha
<4\left(
1+ {\sqrt{6}\over 4\zeta}
\right)
\,.
\end{eqnarray}
For fixed points FP2$_\pm$, the constraint 
for existence is a little more
complicated, as we find;
\begin{eqnarray}
\alpha>3\gamma\left[
1-{2-\gamma\over 2(4-3\gamma)}{1\over \zeta^2}
\right]
&~~~{\rm for}& \gamma<{4\over 3}
\\
{\rm any~values~of} ~\zeta~{\rm and}~\alpha~~~~~~~~~
&~~~{\rm for}& \gamma={4\over 3}
\\
\alpha<3\gamma\left[
1+{2-\gamma\over 2(3\gamma-4)}{1\over \zeta^2}
\right]
&~~~{\rm for}& \gamma>{4\over 3}
\,.
\end{eqnarray}

Next we analyze the solutions of these fixed points.
In what follows, we consider only the expanding universe
($y_F>0$),
i.e.  FP1$_+$, while  FP2$_+$ for $\alpha<3\gamma$
and FP2$_-$ for $\alpha>3\gamma$.

For FP1$_+$, we find
\begin{eqnarray}
{\sigma}&=&{\zeta(4-\alpha)\over \sqrt{2(
6-\zeta^2(4-\alpha)^2)}}\,\tau_*+{\sigma}_0
\\
a_*&=&a_{*0} \exp \left[{\tau_* \over \sqrt{2(
6-\zeta^2(4-\alpha)^2)}}
\right]
\,.
\end{eqnarray}
The cosmic time $t_*$ is given by
\begin{eqnarray}
t_*=t_{*0}\,
\exp \left[{\zeta^2(4-\alpha)^2\,\tau_* \over \sqrt{
2(6-\zeta^2(4-\alpha)^2)}}
\right]
\,.
\end{eqnarray}
Hence the solution of FP1$_+$ is described by
\begin{eqnarray}
{\sigma}&=&{2 \over \zeta(4-\alpha)}\,
\ln\left({t_*\over t_{*0}}\right)+{\sigma}_0
\\
a_*&=&a_{*0} \left({t_*\over t_{*0}}\right)^{p_*}
\,,
\end{eqnarray}
where
\begin{eqnarray}
p_*={2\over \zeta^2(4-\alpha)^2}
\,.
\end{eqnarray}
In this case, matter does not contribute to  the expansion of
the universe, i.e. at the fixed point, $\rho_*=0$.
The cosmic expansion becomes inflationary
if 
\begin{eqnarray}
4\left(
1-{\sqrt{2}\over 4\zeta}
\right)
<\alpha <
4\left(1+{\sqrt{2}\over 4\zeta}
\right)
\,.
\end{eqnarray}


For FP2$_\pm$,
we have
\begin{eqnarray}
{\sigma}&=&\pm {3\gamma \,\tau_*
\over \sqrt{6[3\gamma(2-\gamma)-2(4-3\gamma)(3\gamma
-\alpha)\zeta^2
]}}+{\sigma}_0
\nonumber \\
&&~
\\
a_*&=&a_{*0} \exp \left[{\pm (3\gamma-\alpha)\,\zeta \,\tau_*
\over \sqrt{6[3\gamma(2-\gamma)-2(4-3\gamma)(3\gamma
-\alpha)\zeta^2
]}}
\right]
\,.\nonumber \\
&&~
\end{eqnarray}
The cosmic time $t_*$ is given by
\begin{eqnarray}
t_*=t_{*0}\,
\exp \left[{\pm 3\gamma(4\zeta-\alpha) \,\tau_*
\over 2\sqrt{6[3\gamma(2-\gamma)-2(4-3\gamma)(3\gamma
-\alpha)\zeta^2
]}}
\right]
\,.
\nonumber \\
~
\end{eqnarray}
Hence the solution of FP2$_+$ is described by
\begin{eqnarray}
{\sigma}&=&{2 \over \zeta(4-\alpha)}\,
\ln\left({t_*\over t_{*0}}\right)+{\sigma}_0
\nonumber \\
~\\
a_*&=&a_{*0} \left({t_*\over t_{*0}}\right)^{p_*}
\,,
\end{eqnarray}
where
\begin{eqnarray}
p_*={2(3\gamma-\alpha)
\over 3\gamma(4-\alpha)}
={2\over 3\gamma}+\Delta p_*
\,,
\label{exp_1}
\end{eqnarray}
where 
\begin{eqnarray}
\Delta p_*={2(3\gamma-4)
\over 3\gamma(4-\alpha)}
\,.
\label{delta_p}
\end{eqnarray}
which  describes the deviation from conventional
power exponent with the adiabatic index $\gamma$.
It is due to the
 interaction between the matter fluid 
and the scalar field $\phi$.  It is precisely this interaction that
keeps the ratio  ${\rho_*/ V_*}$ constant at FP2$_\pm$.
The value is obtained from Eq. (\ref{enery_density});
\begin{eqnarray}
&& \frac{1}{6}\left({\rho_*\over V_*}
\right)_{\rm FP2}
=2\zeta^2 \left(y_2^{(\pm)}\right)^2-{
\left(x_2^{(\pm)}\right)^2\over 3}-{1\over 6}
\nonumber \\[.5em]
&&=
{2\left[(4-\alpha)^2+(3\gamma-4)(4-\alpha)
-3\gamma/\zeta^2
\right]\over 2(3\gamma-4)^2+2(3\gamma-4)(4-\alpha)
+3\gamma(2-\gamma)/\zeta^2}
\,,
\nonumber \\
~~
\end{eqnarray}
which is consistent with (\ref{em_cons3}) for the energy density.

In order for this energy density to be positive, we have to impose 
the following condition:
$$
\alpha<
{1\over 2}\left[
(3\gamma+4)-\sqrt{(3\gamma-4)^2+12\gamma/\zeta^2}\right]~~~~{\rm for}~~~
{\rm FP2}_+
\,
$$
$$
\alpha>{1\over 2}\left[
(3\gamma+4)+\sqrt{(3\gamma-4)^2+12\gamma/\zeta^2}\right]
~~~~{\rm for} ~~~
{\rm FP2}_-
\,.
$$

We also find a simple result $p_*=1/2$ follows either by $\alpha=0$ for any
$\gamma$ or by $\gamma=4/3$ for any $\alpha$. This result may be
described by
\begin{eqnarray}
p_*=\frac{1}{2}+\Delta p_*',
\end{eqnarray}
where
\begin{eqnarray}
\Delta p_*'=\frac{(3\gamma -4)\alpha}{6\gamma (4-\alpha)} 
=\frac{\alpha}{4}\Delta p_*
\,.
\end{eqnarray}

We are now looking into more details of the power-law inflation.  We find a power-law inflation by  $\sigma$ if the power exponent of the potential 
$V_*$ is sufficiently small, i.e. $|\zeta(4-\alpha)|<\sqrt{2}$, and if there is no coupling between the matter fluid and 
the scalar field $\sigma$ in the Einstein frame. 
This type of inflation is realized in the fixed point FP1$_+$
even when we include the coupling with the matter fluid, which has nothing to do with  inflation at FP1$_+$.
However, the power-law inflation may occur also for FP2$_\pm$, in which
the coupling to the matter fluid is non-negligible.
From the condition $p_*>1$ for  
the power exponent of the scale factor in the Einstein frame,
we find the following conditions: 
\begin{eqnarray}
\alpha<{6\gamma\over 3\gamma-2}
~~~{\rm or}~~~~\alpha/2>2~~
&~~{\rm for}~~&\gamma<{2\over 3}
\\
\alpha>4~~~~~~~~~~~~~~~~~&~~{\rm for}~~
&\gamma={2\over 3}
\\
4<\alpha<{6\gamma\over 3\gamma-2}~~&~~{\rm for}~~
&{2\over 3}<\gamma<{4\over 3}
\\
{\rm no~case}~~~~~~~~~~~~~~~~~&~~{\rm for}~~
&\gamma={4\over 3}
\\
{6\gamma\over 3\gamma-2}<\alpha<4~~
&~~{\rm for}~~
&\gamma>{4\over 3}
\,.
\end{eqnarray}

This is a new type of inflation.
The potential itself is too steep to  cause inflation,
but the matter fluid assist to cause a faster expansion
because of its coupling to the scalar field. 
Note that including the matter fluid with $\gamma<2/3$ means that we can
have inflation just by this fluid.  But the coupling with the scalar
field does not assist such a type of inflation.  Rather it will restrict the
possibility of inflationary expansion.

Next we have to analyze the stability of the fixed points.
We can analyze it by  perturbations or by use of the phase diagram,
as shown in the previous section.
Here we give the perturbation analysis.
Inserting the perturbations around the fixed points 
$
x=x_F+\delta x, y=y_F+\delta y$
in the basic equations (\ref{scalar_power3}) and (\ref{Friedmann_power3}),
we find a set of the linear perturbation equations (\ref{perturbation_eq})
with
the components of the matrix being
\begin{eqnarray}
\left\{
\begin{array}{l}
A_{xx}=\left(
3\gamma-\alpha
\right)x_F-3 y_F
\\[1em]
A_{xy}=-3x_F+6\zeta^2(4-3\gamma)y_F\\[1em]
A_{yx}=-{\mbox{\small $2-\gamma$}\over \mbox{\small $2\zeta^2$}}x_F+
{\mbox{\small $(4-\alpha)$}\over \mbox{\small $2$}} y_F\\[1em]
A_{yy}={\mbox{\small $(4-\alpha)$}\over \mbox{\small $2$}}x_F-3\gamma y_F
~~.
\end{array}
\right.
\end{eqnarray}
Setting  $\delta x,\delta y \propto e^{\omega \tau_*}$, 
we find the eigen equation for $\omega$ as
 (\ref{eigen_eq})  
with
\begin{eqnarray}
{\rm Tr} A&=&
(3\gamma+2-3\alpha/2)x_F -3(\gamma+1)y_F
\\
\det A &=&
{1\over 2\zeta^2}\left[
(3\gamma-\alpha)(4-\alpha)\zeta^2-3(2-\alpha)\right]x_F^2
\nonumber \\
&+&
3(\alpha\gamma-10\gamma+8)x_F y_F
\nonumber \\
&+&
3\left[
3\gamma-\zeta^2(4-3\gamma)(4-\alpha)\right]y_F^2
\,.
\end{eqnarray}
\begin{widetext}

\begin{minipage}{6.cm}
\hspace*{-1.5em}
\mbox{}\\[.45em]
\includegraphics[scale=.25]{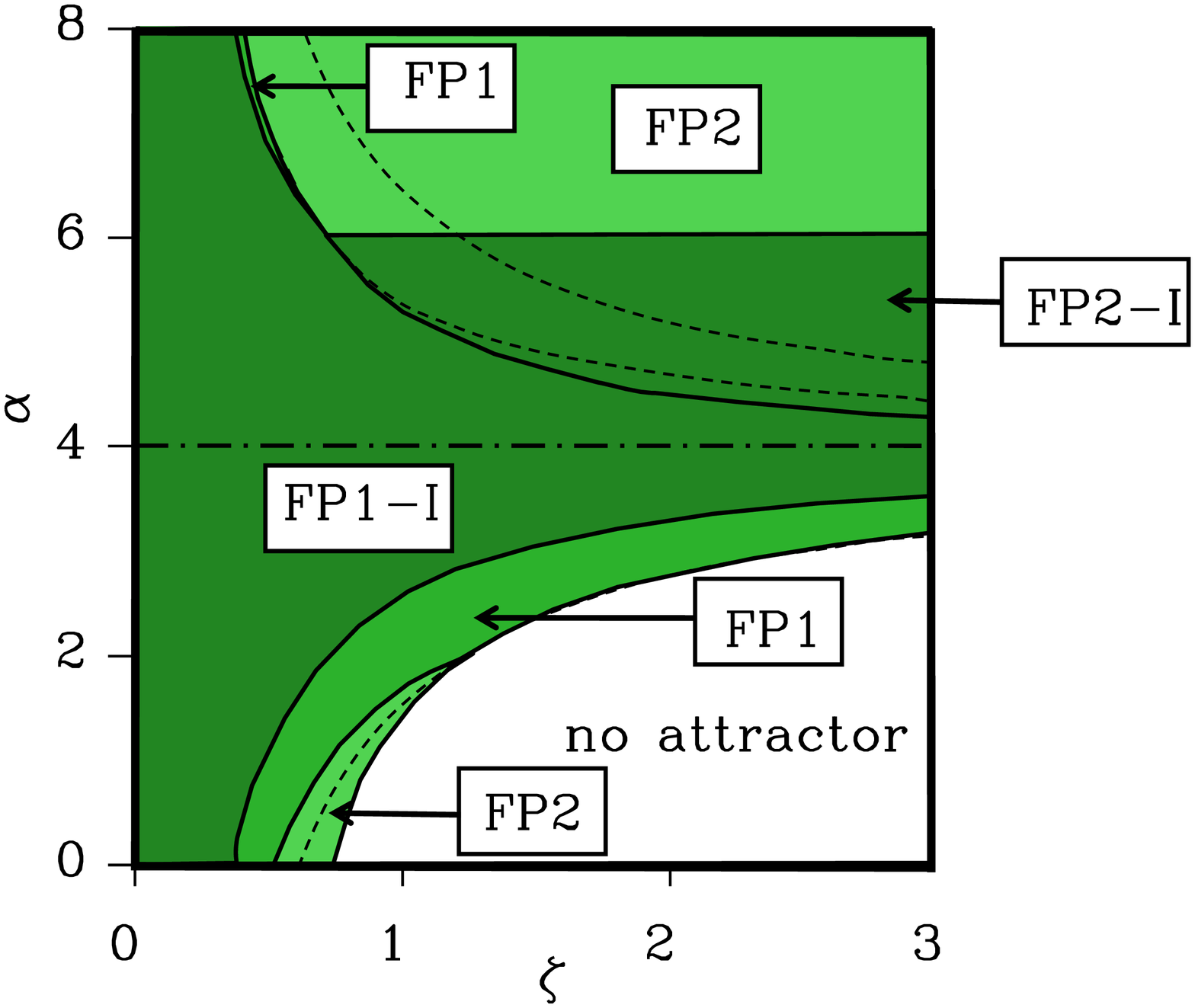}
\\[.86em]
\hspace*{4.1em}
(a) $\gamma=1$ (dust)
\end{minipage}  
\hspace{-1.em}
\begin{minipage}{6.cm} 
\hspace*{-1.5em}
\mbox{}\\[.8em]
\includegraphics[scale=.253]{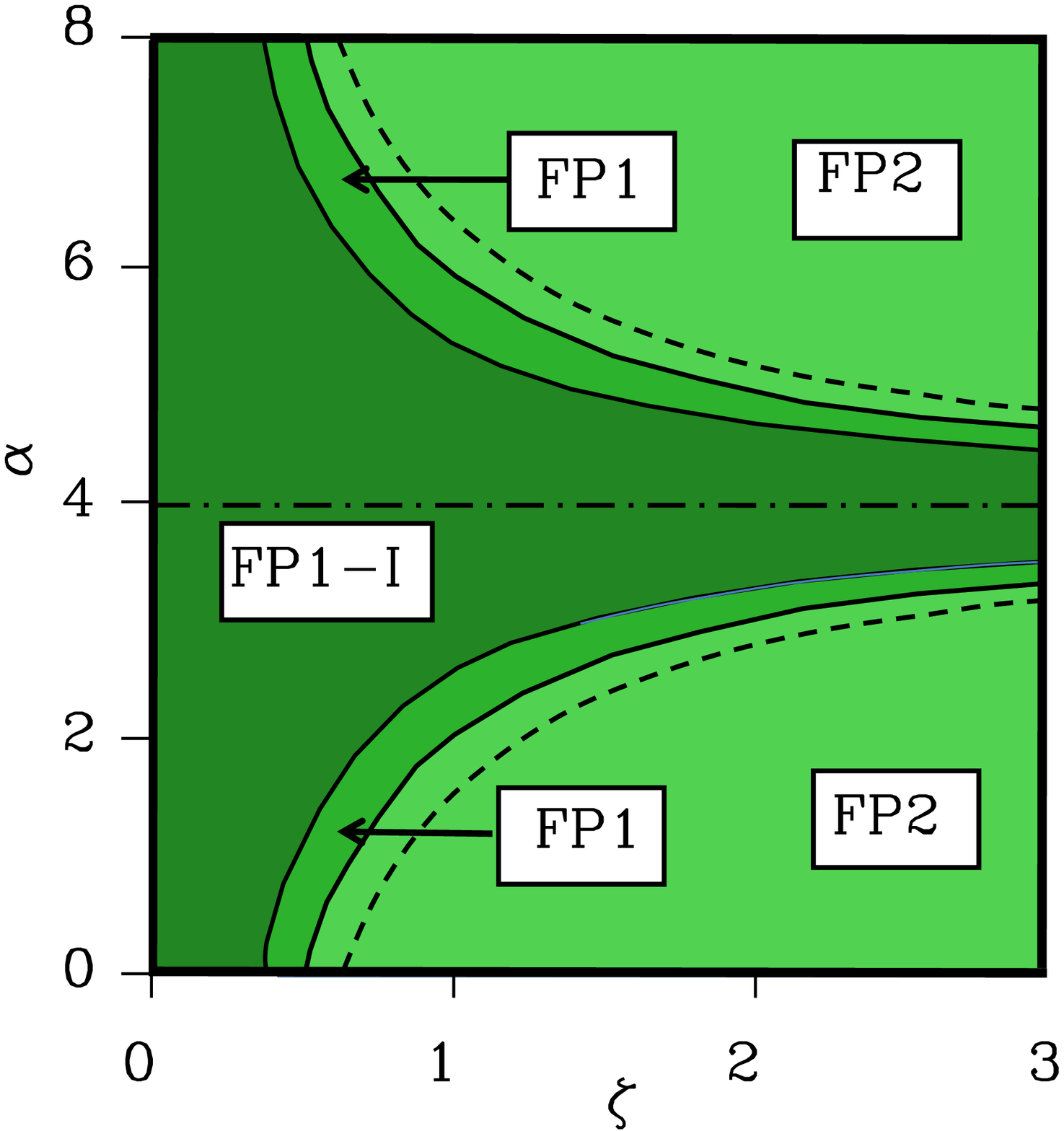}
\\[-.5em]
\hspace*{3.5em}
\mbox{}\\[.15em]
\hspace{2.0em}
(b) $\gamma=4/3$ (radiation)
\end{minipage}  
\hspace{-1.em}
\begin{minipage}{6.cm} 
\hspace*{-1.5em}
\mbox{}\\[.4em]
\includegraphics[scale=.256]{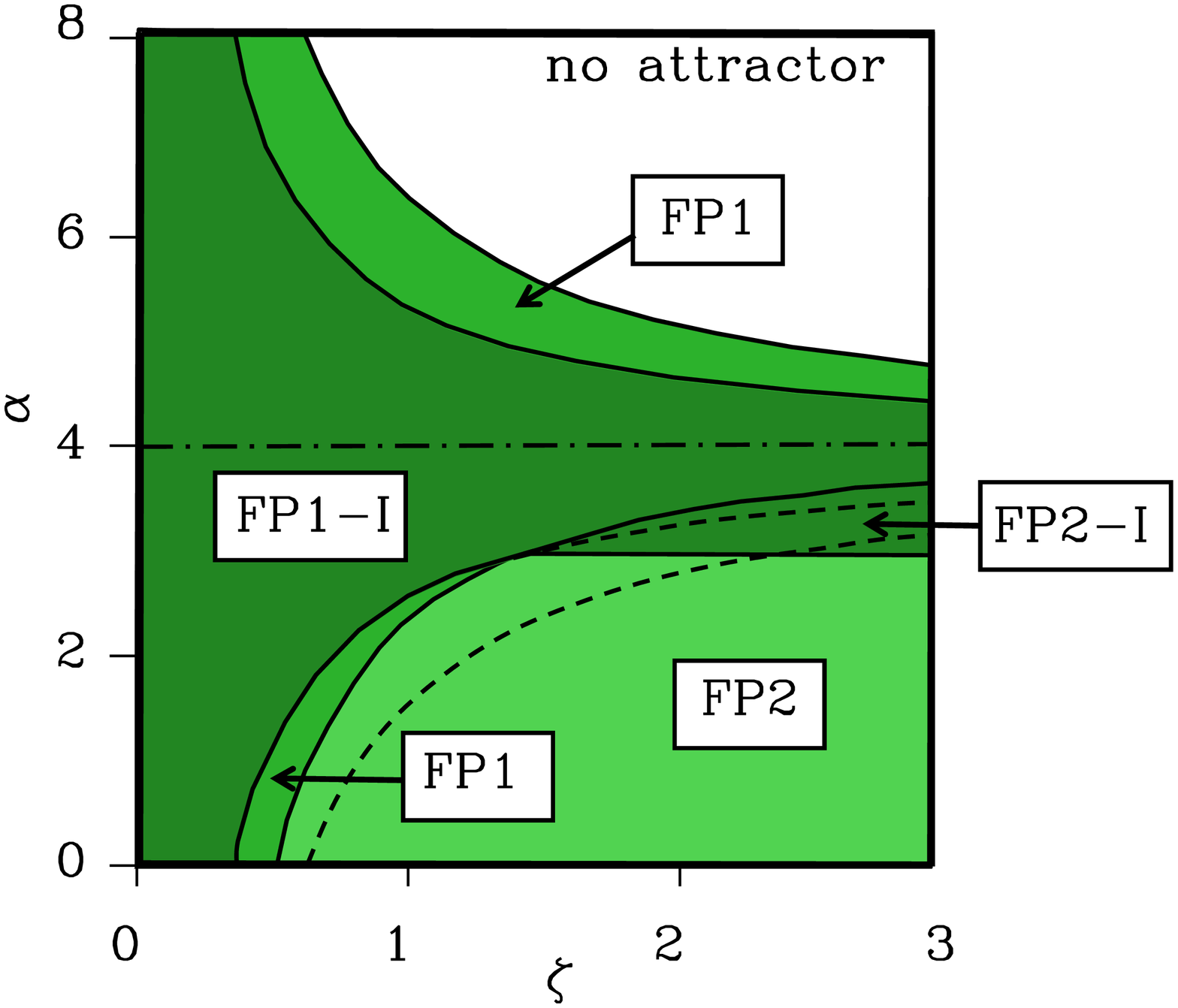}
\\[1.0em]
\hspace*{2.5em}
(c) $\gamma=2$ (stiff matter) 
\end{minipage} 
\mbox{}\\
\begin{figure}[ht]
\caption{\small For various matter fluids
((a) dust [$\gamma=1$], (b) radiation [$\gamma=4/3$], (c) stiff matter
 [$\gamma=2$]), 
we show the ranges in $\zeta$-$\alpha$ plane
of attractor fixed points (FP1$_+$,FP2$_\pm$).
The shaded regions FP1 and FP1-I are those for FP1$_+$,
while FP2 and FP2-I are those for FP2$_\pm$, respectively.
The dark shaded regions FP1-I and FP2-I are the ranges for power-law inflation.There is no attractor solution in the unshaded region.
} 
\label{inflation}
\end{figure}

\end{widetext}

Using these equations, we analyze the stability of
the fixed points  as before.
The result is shown in Fig. \ref{inflation}.
The regions denoted by FP1 and FP1-I give the attractor solution of FP1$_+$,
while those  by FP2 and FP2-I  correspond to the attractors of 
 FP2$_\pm$. 
For FP1$_+$, the inflationary solutions exist in the case of  a flat potential
($\zeta |4-\alpha| <1/2$), but the new type of inflationary solution also appears for FP2$_\pm$ if 
$\gamma\neq 4/3$.
Each region for inflation is also shown by FP1-I or FP2-I.
This new type of inflation is assisted by a coupling between the scalar field 
$\sigma$ and the matter fluid.

In the Jordan frame,
we show only the behavior of the fixed points
FP1$_\pm$ and FP2$_\pm$.
After a conformal transformation, we find them:
\begin{eqnarray}
a &\propto&
  t^{p}
\\
{\phi} &\propto&
  t^{q}
\,,
\end{eqnarray}
where
\begin{eqnarray}
&& \left(p, q\right)=\left\{
\begin{array}{lll}
\left({\mbox{\small $2(1-(4-\alpha)\zeta^2)$}
\over \mbox{\small $(2-\alpha)
(4-\alpha)\zeta^2$}},
{\mbox{\small $2$}\over 
\mbox{\small $2-\alpha$}}\right) 
& {\rm for} & {\rm FP1}_\pm
\\
\left(-{\mbox{\small $2\alpha$}
\over \mbox{\small $2-\alpha$}},
{\mbox{\small $2$}
\over \mbox{\small $2-\alpha$}}\right) &
 {\rm for} & {\rm FP2}_\pm
\end{array}
\right.
\nonumber \\
~~
\end{eqnarray}

The scale factor  in the Jordan frame 
 for FP2$_\pm$ is
not constant except for $\alpha=0$.
It is expanding for $\alpha<0$ or 
$\alpha>2$, while
it is contracting for $0<\alpha<2$.
For $\alpha=2$, we find $a\propto
\exp[-{2\over 3\gamma}t]$.

\section{Concluding Remarks}

We have presented a formulation in which we trace the temporal development of
cosmological solutions of the scalar-tensor theory in  two-dimensional phase
space.  Thanks to assuming a simple equation of state, 
in Sections III, IV and
V, we have obtained two different sets of fixed points, FP1$_\pm$ and
FP2$_\pm$.   Conditions of obtaining attractors are studied in detail.  

We have established the attractor nature of  the fixed points.
At FP2$_\pm$, the scale factor behaving like a constant
and $t_*^{1/2}$ in the Jordan and the Einstein frames, respectively,
for $\zeta^2>1/4$ or $\xi <1/2$ ($\omega<-1/2$) with $\epsilon =-1$, 
when  we have a simple cosmological constant in the Jordan frame.  
This solution is also accompanied with the proportionality between
$\rho_*$ and $V_*$, called a scaling behavior, which is going to be
replaced by the ``interlacing'' behavior, as exemplified in Fig. 5.8. of 
\cite{Fujii_Maeda}, by further extending the model.

An extension to the power-law potential in Section V has shown 
 that  the coupling
between the scalar field and the matter provides a new type of inflation with
$\gamma$ off the conventional choice $4/3$,
even if the potential of the scalar field 
is too steep to cause inflationary expansion by itself.

We have also learned that, even confining ourselves to the fixed-point
solutions, reaching an attractor can be somewhat complicated if we have
another fixed point, as we faced in the example of 
Fig. \ref{f2}, though a simple recipe is shown to be applied by adopting
either of the two attitudes suggested; limiting the range of
$\zeta^2$ or being content with an {\em a posteriori} consideration that
initial values of $x$ and $ y$ had happened to be in favor of reaching
the attractor.  The first appears to 
 be the case in our realistic choice ($\zeta =1.5823$) to fit the observed
 accelerating universe shown in Fig. 5.8 of \cite{Fujii_Maeda}, for
 example.  The  second of the above suggestions might also apply to
 drifting to infinities, as will be demonstrated in Appendix B.

If there is a spatial curvature, the result will be changed.
However, we can show that the fixed point is still 
an attractor if the universe expands very fast, i.e,
if it is an inflationary universe, as shown in Appendix A.

The most intriguing result in the present scalar-tensor theory is,
however, that  the {\em static} universe in the Jordan
frame is an attractor, an unavoidable fate in the presence of
the cosmological constant.  There is no smooth limit as $V_0
\rightarrow 0$.  As we re-iterate, the Jordan frame features
truly constant masses of microscopic fields, according to the
Brans-Dicke model, originally intended to qualify this frame to be 
physical, allowing a non-static universe in the absence of the
cosmological constant.  Its presence alters the entire situation,
forcing us to accept  $ma ={\rm const}$.  This entails  eventually that
the universe in the Einstein frame expands in the {\em same} rate as
the (time-dependent) microscopic  length standard.  This {\em crisis}
 will be evaded only by leaving the Brans-Dicke model, as was elaborated in 
\cite{Fujii_Maeda},\cite{ptpinv} together with the ensuing consequences.  The view that  this crisis hinges upon the attractor nature of the solutions
is now reinforced even more strongly by our study in this article.

We also emphasize that the extension 
to the power-law potential leaves the
 above crisis unsolved. 
The exponent $1/2$ in radiation-dominance in the Einstein frame is a unique
consequence of $\gamma =4/3$ independent of the way $V_0$ is modified by
the scalar field.   The argument on
 the power-law inflation also remains unaffected, as we point out, by the
 structure of the mass term.

\acknowledgments

KM would like to thank DAMTP and the Centre for Theoretical Cosmology,
Cambridge University
for hospitality during his stay in September, 2008.
YF thanks Takatoshi Ichikawa for his help in preparing Fig. 6,
and useful advices for computing in Fig. 7.
This work was partially supported 
by the Grant-in-Aid for Scientific Research
Fund of the JSPS (No.19540308) and for the
Japan-U.K. Research Cooperative Program,
and by the Waseda University Grants for Special Research Projects. 



\appendix
\newpage

\section{The Effect of Curvature Term ($k\neq 0)$}
\label{curvature_effect}
In this Appendix, we study the curvature effect.
It may be convenient to analyze the equations 
rewritten by new variables because the fixed points 
are constant.
 We discuss the cases with a cosmological constant
and with  a power-law potential separately.
\subsection{The case with a cosmological constant}
 We show the curvature effects both in the Einstein 
and in the Jordan frames in this order.
\subsubsection{Curvature term in the Einstein frame}

The curvature term is proportional to 
$e^{4\zeta{\sigma}}/ a_*^2$.
We evaluate the time evolution of this term by
\begin{eqnarray}
&&{d\ln (e^{4\zeta{\sigma}}/ a_*^2)\over d\tau_*}
=2(2x-y)
\end{eqnarray}
We find the behavior near the fixed point $(x_F,y_F)$ as
\begin{eqnarray}
&&{d\ln (e^{4\zeta{\sigma}}/ a_*^2)\over d\tau_*}
=2(2x_F-y_F)
\nonumber \\
&&~~~~~~=
\left\{
\begin{array}{lll}
{\mbox{\small $8\zeta^2-1$}\over \mbox{\small \raisebox{-.5em}{$2\zeta\sqrt{3-8\zeta^2}$}}}
&{\rm for}&{\rm FP1}_+\\[1em]
{\mbox{\small $\sqrt{\gamma}$}
\over \mbox{\small  \raisebox{-.5em}{$\sqrt{2(2-\gamma-2(4-3\gamma)\zeta^2)}$}}}
~~~&{\rm for}&{\rm FP2}_+~~~
\end{array}
\right.
\label{curvature_effect}
\end{eqnarray}
Hence we find that 
the curvature term decreases in time 
for inflationary solution 
at FP1$_+$ ($\zeta^2<1/8$).
It can be ignored.
However, it will grow in time if the universe expands
without acceleration. It will becomes important as
the same as the usual case.
\subsubsection{Curvature term in the Jordan frame}
\label{curvature_Jordan}
The time evolution of the curvature term 
in the Jordan frame
is given by
\begin{eqnarray}
{d\ln (\phi^2/a^2)\over d\tau}
=2(\Phi'-\phi{\cal H})
=2(2\Phi'-{\cal H})
=2(2x-y)\,.
\nonumber \\
~
\end{eqnarray}
We find the behavior near the fixed point $(x_F,y_F)$
as
\begin{eqnarray}
{d\ln ({\phi}^2/a^2)\over d\tau}
=2(2x_F-y_F)\,,
\end{eqnarray}
which is exactly the same as Eq. 
(\ref{curvature_effect}).
Hence the curvature term near the fixed point FP1$_+$ 
 is not important for inflationary solution 
($\zeta^2<1/8$).

\subsection{The case with a power-law potential}

In this case, we can repeat the same analysis.
The curvature term is proportional to 
$e^{(4-\alpha){\zeta}{\sigma}}/ a_*^2$.
So the time evolution near the fixed point is given by
\begin{eqnarray}
&&{d\ln (e^{(4-\alpha){\zeta}{\sigma}}/ a_*^2)\over d\tau_*}
=\zeta\left[2(2x_F-y_F)-\alpha x_F\right]
\nonumber \\
&&~~~~=
\left\{
\begin{array}{lll}
{\mbox{\small $2\zeta^2(2-\alpha/2)^2-1$}
\over \mbox{\small \raisebox{-.5em}{$\zeta\sqrt{2(6-\zeta^2(\alpha-4\zeta)^2}$}}}
&{\rm for}&{\rm FP1}_+\\[1em]
{\mbox{\small $\zeta[6\gamma-\alpha(3\gamma-2)]$}
\over \mbox{\small \raisebox{-.5em}{$\sqrt{6[3\gamma(2-\gamma)-2(4-3\gamma)(3\gamma
-\alpha)\zeta^2
]}$}}}
&{\rm for}&{\rm FP2}_+
\end{array}
\right.
\nonumber \\
&&~~
\end{eqnarray}
The curvature term is not important for inflationary solution 
near the fixed point  FP1$_+$ ($\zeta^2(4-\alpha)^2<2$)
and near the fixed point
 FP2$_+$ ($\alpha>6\gamma/(3\gamma-2)$ with $\gamma>4/3$)
or  FP2$_-$ ($\alpha<6\gamma/(3\gamma-2)$ with $2/3<\gamma<4/3$).
\\[1em]

\section{Additional solutions for radiation-dominance in the Jordan frame}
\label{extra_solution}

We often relied on the numerical approach to solve the cosmological equations
in the Jordan frame because the solution is characterized by the simplest
aspect of the static universe.    We encounter,
however, another complicated aspect to be discussed in what follows.

We started conveniently from
\begin{eqnarray}
6\varphi H^2 &=& -\frac{1}{2} \dot{\phi}^2 +V_0 +\rho -6H\dot{\varphi},
\label{attr_1}\\
\ddot{\varphi}+3H\dot{\varphi}&=&4\zeta^2 V_0 ,
\label{attr_2}\\
\dot{\rho}+4H\rho &=&0,
\label{attr_3}
\end{eqnarray}
where $\varphi =(\xi/2)\phi^2$ in terms of which (\ref{scalar_ST}) has
been put into a simplified form in (\ref{attr_2}), as in 
\cite{Fujii_Maeda}.  We also write $V_0=\Lambda =1$.

Since $H$ occurs always without derivative,
we may eliminate it by using (\ref{attr_3}), for example; 
\begin{eqnarray}
H=-\frac{1}{4}\frac{\dot{\rho}}{\rho},
\label{rad3_11}
\end{eqnarray}
Eqs. (\ref{attr_1}) and  (\ref{attr_2}) are then
put into 
\begin{eqnarray}
&& 3\left(\frac{\dot{\rho}}{\rho}\right)^2\varphi 
-12\frac{\dot{\rho}}{\rho}\dot{\varphi}+2\xi^{-1}
\frac{\dot{\varphi}^2}{\varphi}=8\left( V_0 +\rho \right),
\label{rad3_13}\\
&& \ddot{\varphi}-\frac{3}{4}\frac{\dot{\rho}}{\rho}
\dot{\varphi}=4\zeta^2 V_0,
\label{rad3_14}
\end{eqnarray}
which are to be solved by giving three initial values of $\rho, \varphi,
\dot{\varphi}$.

We notice, however, that we solve (\ref{rad3_13}) and
(\ref{rad3_14}) first with respect to $\dot{\rho}/\rho$.  This involves
solving an {\em algebraically}  quadratic equation for $\dot{\rho}/\rho$,
thus producing two {\em differential} equations, hence  resulting in two 
separate solutions.  An example of
numerical solutions is shown in Figs. $\ref{f3} (a) {\rm and} (b)$.  In
Fig. \ref{f3} (a), developed basically from Fig. 4.1 of \cite{Fujii_Maeda}, we
do find asymptotic behaviors $H\rightarrow 0, \dot{\phi}\rightarrow
\sqrt{4 V_0/(6\xi -1)}, \rho \rightarrow -3V_0 (2\xi -1)/(6\xi
-1)$ corresponding to an attractor solution.  Fig. \ref{f3}(b) illustrates,
on the other hand, the solution of another equation, but sharing the same
initial values of $\rho, \varphi$ and $\dot{\varphi}$ as discussed in
(a).  This one represents, 
however,  a shrinking universe taking place in a short time.  This type of the
second solution occurs nearly always.  It even appears as if we are
going to lose an opportunity to reach the fixed-point attractor.

\begin{widetext}
\mbox{}\\[-10.0em]
\hspace*{-1.5em}
\begin{minipage}{6.cm}
\hspace*{-1.0em}
\includegraphics[keepaspectratio,width=9.4cm]{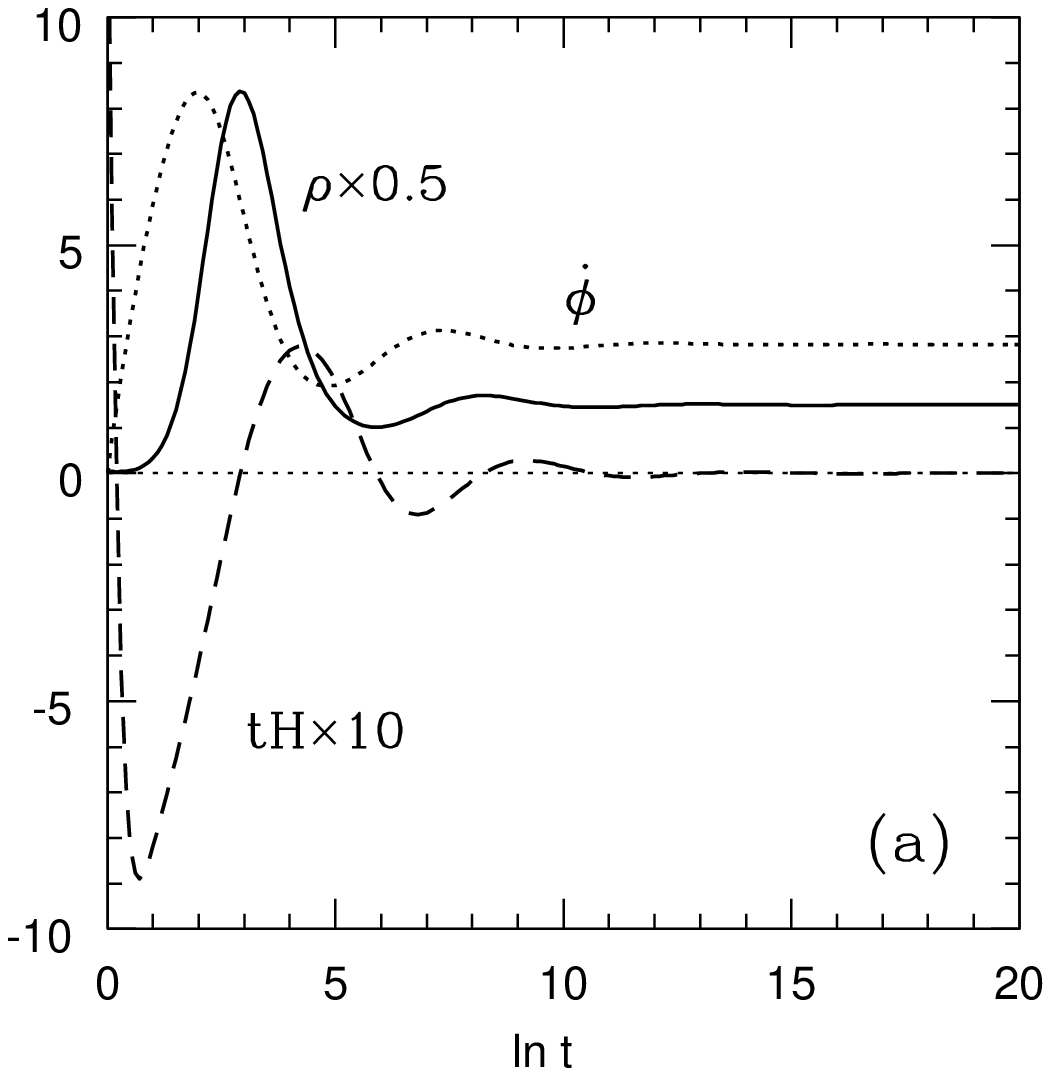}
\end{minipage}
\hspace{-.6em}
\begin{minipage}{6.cm}
\hspace*{-1.0em}
\includegraphics[keepaspectratio,width=9.4cm]{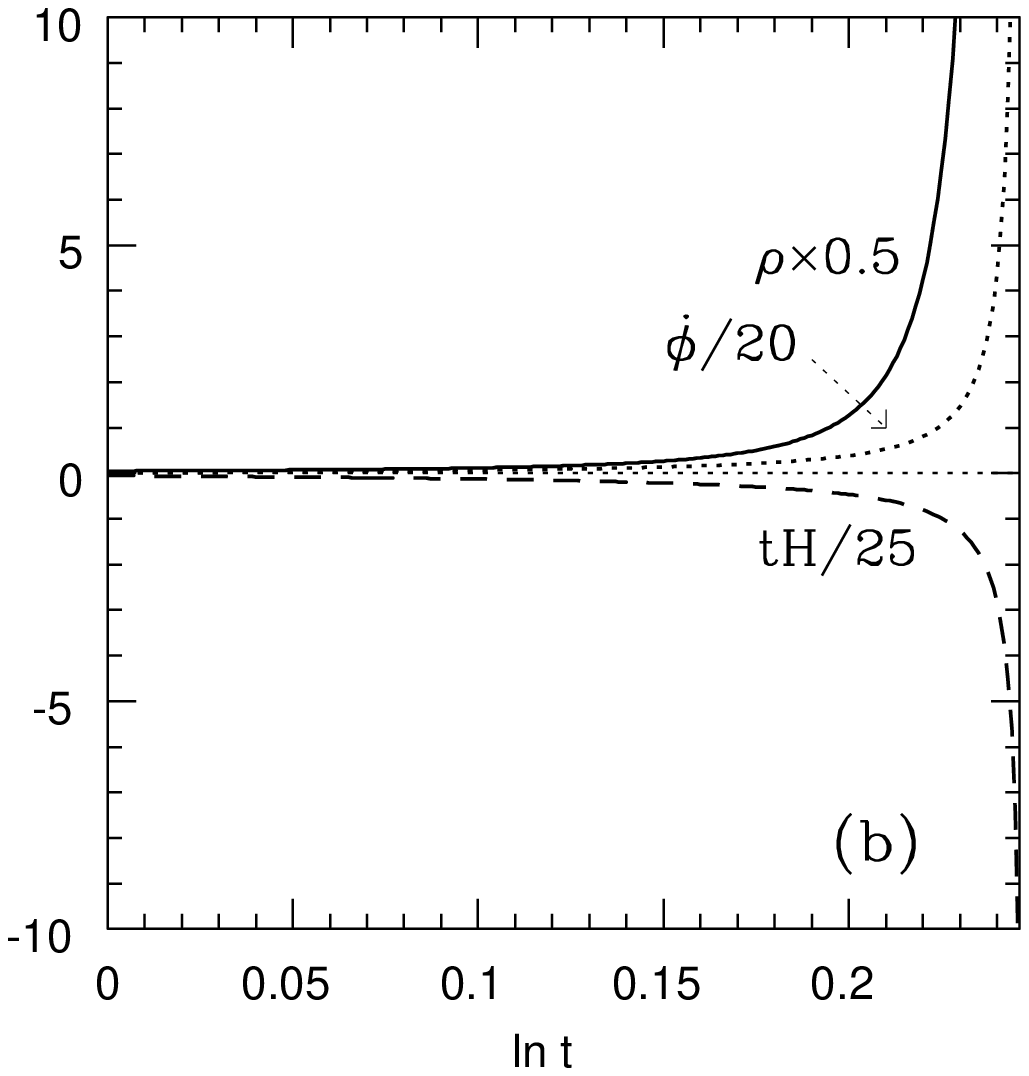}
\end{minipage}
\hspace{-.6em}
\begin{minipage}{6.cm}
\hspace*{-1.0em}
\includegraphics[keepaspectratio,width=9.4cm]{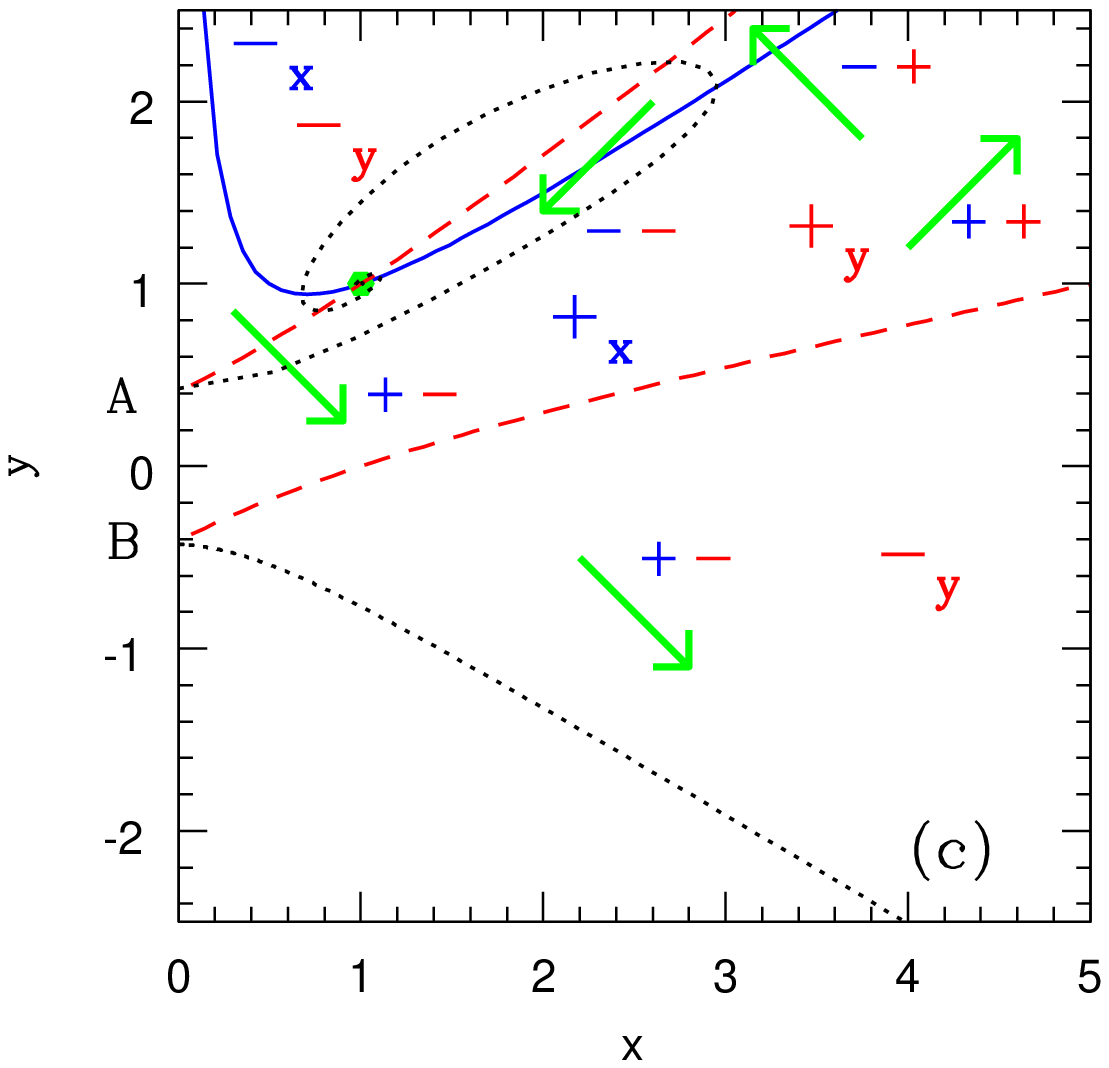}
\end{minipage}
\mbox{}\\[-3.0em]
\begin{figure}[h]
\caption{Two solutions for the initial values at $\ln t=0$; $\rho =0.1,
 \varphi =0.1, \dot{\varphi}=0$ together with $V_0=1$.  The asymptotic
 values mentioned above, 
 hence (\ref{sigmardsol})-(\ref{scalerdsol}) are reached in (a), while these
 quantities in (b) diverge quickly at $\ln t\rightarrow 0.248$, in this
 example.  Behaviors of the solutions shown in (a) and (b) are
 represented by two different trajectories  in phase space of $x, y$ of
 (c), starting at two different points $A$ and $B$.  The upper curve 
 corresponding to (a) spirals finally down into the attractor denoted by
 a blob (green) at $x=y=1$, while the lower curve for (b) runs steadily
 toward the right-lower infinity, as ``guided'' by the arrow.
}
\label{f3}
\end{figure}
\mbox{}\\[-2.0em]

\end{widetext}

Fortunately, as it turns out however, the phase-space description is so
generous that the resulting two solutions are accommodated as those due
to two different initial locations of $y$, as demonstrated in Fig. \ref{f3}(c).  Again
for a typical solution for $\xi=1/4$, we find two trajectories for the
solutions of (\ref{rad3_13})-(\ref{rad3_14}).  Both start at $x=0$
corresponding to our initial value $\varphi'=0$, but with $y$ separated
from $y=0$ nearly the same distance in the opposite directions.  The
upper curve starts at the point $A$.   After a loop-like trip, it
finally spirals down into the fixed point at $x=y=1$, as usual.  Another curve
starting at $B$ drifts steadily toward $x\rightarrow \infty,
y\rightarrow -\infty$.  Both trajectories started immediately ``outside"
the $y'=0$ boundaries.

As we find from $y=\zeta^{-1}{\cal H}_*$  preceding (\ref{SA1}) and
(\ref{SA2}), the solutions with $y>0$ and $y<0$ correspond to the expanding
and contracting universes, respectively, in the Einstein frame.  From
this point of view, what we are in the expanding universe at quite a
late epoch must be a consequence of the initial value selected to be $y>0$, as
represented by the upper trajectory in Fig. \ref{f3} (c).  In this sense 
the solution like  Fig. \ref{f3} (b) is excluded leaving the attractor
solution alone.

\section{ Dust-dominance  in the scale-invariant model }
\label{dust-universe-scale-inv}

In sections III and IV we discussed cosmological solutions mainly in the
radiation-dominated universe finding a crisis arising from too much
time-dependent masses of particles evaded finally by departing from the
Brans-Dicke model, even at the risk of WEP violation.  The same type of
analysis of dust-dominance suffers more seriously because it
entails $a_*\sim t_*^{1/2}$, as shown by (\ref{scalerdsol}) even for $\gamma =1$. 
As was discussed in \cite{Fujii_Maeda}, the remedy comes simultaneously from the
scale-invariant model intended to overcome the crisis for
radiation-dominance.  We sketch below how this model provides attractor
solutions also for the dust-dominated universe.  See  Chapter 4.4.3 of  
\cite{Fujii_Maeda}  and  Section 3.4 of \cite{ptpinv} for more details.


The field equations for $\sigma$ and $\rho$ turn
 out to be given by (\ref{fundeq2_2}) and (\ref{fundeq2_3}) with the right-hand
 sides removed to the classical approximation.

As a remarkable difference from
 (\ref{scalerdsol}) we find
\begin{eqnarray} 
a_*&=&a_{*0} \left(t_*/ t_{*0}\right)^{2/ 3}
\,,
\label{scalerdsolrv}
\end{eqnarray}
in agreement with the conventional law of expansion.
 Eqs. (\ref{rad3_5}) and (\ref{rad3_6}) are replaced by 
\begin{eqnarray}
&& 
 x'=\zeta \left( 2x^2 -3xy +1 \right)
\,,
\label{rvx} \\
&&  
y'=\frac{1}{ 2\zeta}\left( -\frac{1}{2}x^2 +\frac{1}{4} +\zeta^2
			 (4x-3y)y \right)
\,,
\label{rvy}
\end{eqnarray}
respectively.  The solutions with (\ref{sigmardsol}),
(\ref{scalerdsolrv}) and 
\begin{eqnarray}
\exp\left(-4\zeta\sigma_0\right)=\frac{1}{16}\zeta^{-2},
\end{eqnarray}
in place of (\ref{sigmabar1}) are obtained for $x=1/\sqrt{2},
y=2\sqrt{2}/3$, corresponding to an attractor yielding $x'=y'=0$.

A similar distinction between the elliptic and the hyperbolic curves as
 in radiation-dominance occurs also for $\zeta^2 <3/8$ and  $\zeta^2
 >3/8$, respectively.  The same recipe should apply as mentioned toward
 the end of section VI.

We add that the scale invariance coming from the absence of
dimensional coupling constants has an advantage that $\sigma$ serves as
a massless Nambu-Goldstone boson which will acquire a small mass after
the invariance is finally broken explicitly through loops, as
discussed in Section 6.3 of \cite{Fujii_Maeda}.

\section{Another approach to the power-law potential}

It seems also useful to apply (\ref{attr_1})-(\ref{attr_3}) to the
power-law potential to offer a simplified alternative to derive the same
common result $p_*=1/2$ for radiation-dominance as stated immediately
after in (\ref{exp_1}).

We multiply $V_0$ in (\ref{attr_1}) and (\ref{attr_3}) by
$\phi^\alpha$.  We search for the solution of the type
\begin{eqnarray}
a(t)\sim t^p,
\label{scalef}
\end{eqnarray}
and 
\begin{eqnarray}
\phi(t)\sim t^\beta.
\label{varphiff}
\end{eqnarray}

By substituting them into (\ref{attr_2}) modified as above  and
comparing the exponents of $t$ we obtain
\begin{eqnarray}
\beta =\frac{2}{2-\alpha},
\label{mu1}
\end{eqnarray}
implying that $\alpha =0$ corresponds to $\beta =1$.  In the modified (\ref{attr_1}), on the other hand, we find all the
terms other than $\rho$ to behave like $t^{2\beta-2}$, while (\ref{attr_3})
entails $\rho \sim t^{-4p}$.  For a consistent approach we expect $2\beta
-2=-4p$, or
\begin{eqnarray}
p=\frac{1-\beta}{2}=-\frac{\alpha}{2(2-\alpha)}.
\label{mu2}
\end{eqnarray}
This point was not properly recognized when
 it was erroneously stated in Appendix B of \cite{ptpinv} that only $\alpha
 =0$ is consistent with the  physically acceptable condition $p_*
 =1/2$ (The coefficients $\alpha$ and $\beta$ in Appendix B of 
 \cite{ptpinv}  are 
 replaced by $\alpha/2$ and $\beta/2$, respectively, according to the
 present notation.).

Now from (\ref{change_time_coordinate}) and $\Omega \sim \phi$
combined with (\ref{varphiff}),  we obtain
\begin{eqnarray}
dt_*=\Omega dt\sim t^{\beta}dt,
\label{ttotstar1}
\end{eqnarray}
which is integrated to give
\begin{eqnarray}
t_* \sim t^{\beta+1},
\label{ttotstar2}
\end{eqnarray}
where we have ignored inessential coefficients for simplicity.

In the same context we also use (\ref{change_scale_factor}) to derive
\begin{eqnarray}
a_* =\Omega a\sim t^{\beta +p}. 
\label{astar1}
\end{eqnarray}
Combining this with (\ref{mu2}) and (\ref{ttotstar2})  we identify the
right-hand side with
$t_*^{p_*}$ where
\begin{eqnarray}
p_*=\frac{1}{\beta +1}\left( \beta+p \right) =\frac{1}{2},
\label{astara}
\end{eqnarray}
which  turns out to be the same as the result for the purely constant
$V_0 =\Lambda$, in agreement with (\ref{exp_1}) with $\alpha =0$ for any
$\gamma$.  This justifies that the present solution to be an attractor.

We also notice that the universe is no longer static in the Jordan frame, as
shown in (\ref{mu2}).  We may no longer rely
on the simplest argument $am =a_*m_*={\rm const}$ to leave the BD model.
According to Appendix D  of \cite{ptpinv}, particularly its (3.10) and  a more
general procedure developed there, however, the mass $m_*$ of the matter fields in the Einstein
frame is related to $m$ in the Jordan frame as 
\begin{eqnarray}
m_*=\Omega^{-1}m \sim t^{-\beta}m.
\label{ttotstar3}
\end{eqnarray}
From (\ref{mu1}) and (\ref{ttotstar2}) we find
\begin{eqnarray}
t^{-\beta}\sim t_*^{-\beta/(\beta+1)}\sim t_*^{-(1/2)/(1-\alpha/4)}.
\label{ttotstar4}
\end{eqnarray}
The exponent $-(1/2)/(1-\alpha/4)$ is not exactly the same as $-1/2$
which would have implied that the universe looks static if measured with
respect to the microscopic length standard, but is nevertheless far from zero 
as expected if the Einstein frame is  qualified to be a physical conformal 
frame for any reasonable choice of
$\alpha$.  In this sense the {\em crisis} for the purely constant
$V_0$ as discussed before is not evaded 
by multiplying it by the scalar field.  Departure from the BD model
seems still unavoidable.  We also add that the argument for the constant
$m_*$ with the assumed scale-invariant model remains unaltered by the
multiplied scalar field.


\begin{thebibliography}{99}
\bibitem{footnote1} This relation is
         an outgrowth of the earlier attempt \cite{yf82}.
	For more details, see Chapter 4.4.2. of \cite{Fujii_Maeda}, 
	towards the end of Section 4.1 of \cite{ptpinv}, and Eqs. (16),
	(17) together with the footnote 12 of \cite{antip}.
\end{thebibliography}
\end{document}